\documentclass[doublespacing]{bmcart}

\usepackage{amsthm,amsmath,amstext}

\RequirePackage{natbib}
\usepackage[utf8]{inputenc} 
\usepackage{times}
\usepackage{txfonts}
\usepackage{relsize}
\usepackage{cleveref}
\usepackage[section]{placeins}
\usepackage[switch, mathlines]{lineno}
\usepackage[strings]{underscore}

\newcommand\Tstrut{\rule{0pt}{2.9ex}}         
\newcommand\Bstrut{\rule[-1.2ex]{0pt}{0pt}}   
\newcommand\TBstrut{\Tstrut\Bstrut}           

\def\includegraphics{}

\usepackage{graphicx,multicol}

\startlocaldefs
\endlocaldefs

\begin{document}

\begin{frontmatter}

\begin{fmbox}
\dochead{Research}


\title{DESTINY: Database for the Effects of STellar encounters on dIsks and plaNetary sYstems}


\author[
   addressref={aff1,aff2},                   
   corref={aff1},                       
   email={bhandare@mpia.de}   
]{\inits{A}\fnm{Asmita} \snm{Bhandare}}
\author[
   addressref={aff2,aff3},
   email={s.pfalzner@fz-juelich.de}
]{\inits{S}\fnm{Susanne} \snm{Pfalzner}}


\address[id=aff1]{
  \orgname{Max-Planck-Institut f\"ur Astronomie}, 
  \street{K\"onigstuhl 17},                     %
  \postcode{69177}                                
  \city{Heidelberg},                              
  \cny{Germany}  
  }
\address[id=aff2]{%
  \orgname{Forschungszentrum J\"ulich, J\"ulich Supercomputing Center},
  \street{Wilhelm-Johnen-Strasse},
  \postcode{52428}
  \city{J\"ulich},
  \cny{Germany}
}
\address[id=aff3]{%
  \orgname{Max-Planck-Institut f\"ur Radioastronomie},
  \street{Auf dem H\"ugel 69},
  \postcode{53121}
  \city{Bonn},
  \cny{Germany}
}


\end{fmbox}


\begin{abstractbox}

\begin{abstract} 
Most stars form as part of a stellar group. These young stars are mostly surrounded by a disk from which potentially a planetary system might form. Both, the disk and later on the planetary system, may be affected by the cluster environment due to close fly-bys. The here presented database can be used to determine the gravitational effect of such fly-bys on non-viscous disks and planetary systems. The database contains data for fly-by scenarios spanning mass ratios between the perturber and host star from 0.3 to 50.0, periastron distances from 30 au to 1000 au, orbital inclination from $0^{\circ}$ to ${180^\circ}$ and angle of periastron of $0^{\circ}$, ${45^\circ}$ and ${90^\circ}$. Thus covering a wide parameter space relevant for fly-bys in stellar clusters. The data can either be downloaded to perform one's own diagnostics like for e.g. determining disk size, disk mass, etc. after specific encounters, obtain parameter dependencies or the different particle properties can be visualized interactively. Currently the database is restricted to fly-bys on parabolic orbits, but it will be extended to hyperbolic orbits in the future. All of the data from this extensive parameter study is now publicly available as DESTINY.
\end{abstract}




\end{abstractbox}
%

\end{frontmatter}



\graphicspath{{graphics/}}

\section{MAIN TEXT}
\label{sec:intro}
The discovery of currently $\sim$ 4000 
\footnote{https://exoplanetarchive.ipac.caltech.edu/index.html} 
extrasolar planets shows the ubiquity of planet formation outside our own solar system. Protoplanetary disks provide the basic material required for the formation of such planetary systems. Recent observations of protoplanetary disks surrounding stars in stellar clusters have been milestones in understanding some of the most fundamental questions regarding the formation and evolution of planetary systems, like disk sizes, lifetimes etc. \citep[for example,][]{Moor2013,Mann2014,Bally2015,Tobin2015,Andrews2018}

Currently, most planet formation theories treat planet formation as an isolated event, where the planets form from the disk surrounding a single or binary star \citep[for example,][]{Bromley2011,Baruteau2014,Bitsch2015,Bromley2015}. However, in accordance with the currently accepted star formation scenarios, most young stars are not formed in isolation but as a part of a group of stars commonly referred to as star cluster \citep{Clarke2000,Lada2003,Porras2003}. As these young stars are at least initially surrounded by protoplanetary disks, the cluster environment might have significant effects on the evolution of these disks \citep[for an overview, see][and references therein]{Hollenbach2000,Williams2011}. The prime external processes that influence the evolution and properties of protoplanetary disks are external photoevaporation due to nearby massive stars \citep{Johnstone1998,Adams2004,Font2004,Clarke2007,Dullemond2007,Gorti2009,Owen2010,Owen2012,Rosotti2015}, head-on accretion \citep{Wijnen2017} and gravitational interactions during fly-bys \citep{Heller1995,Hall1996, Clarke1993,PfalznerVogel2005,Kobayashi2001,Kobayashi2005,Breslau2014,Jilkova2016,Bhandare2016,Winter2018a,Winter2018b,Cuello2019}. In our study we focused on the effects of stellar fly-bys covering a wide parameter space. These results are now publicly available in the database DESTINY 
\citep{DESTINY} \footnote{http://www3.mpifr-bonn.mpg.de/encounter-properties/}.

In star cluster environments stellar fly-bys can reduce the disk mass, change the disk's angular momentum, lead to additional accretion or truncate the protoplanetary disk \citep{Rosotti2014,Vincke2015,Vincke2016,Zwart2016,Cai2018,Zwart2019,Vincke2018,Winter2018a,Winter2018b,Concha2019}. Depending on the local stellar density even already formed planetary systems might be influenced by the gravitational interactions with neighbouring stars, even though this is much less frequent \citep{Thies2005,Kobayashi2005,Jilkova2015,Mustill2016,Pfalzner2018a}. Stellar fly-bys can lead to planets becoming unbound and/or planetesimals being launched from the disks. It is hence important to parameterize the disk properties like disk mass, angular momentum, energy, disk size etc. The truncation radius can prove to be useful to constrain the region within which enough matter would be available for planet formation. 

Many of the simulations mentioned above do not give quantitative results of the effect of fly-bys on disks. Some provide either fit formulae or tabulated values of specific disk properties after fly-bys \citep{Kobayashi2005,Olczak2006,ovelar2012,Breslau2014,Bhandare2016,Cuello2019}. Although this is valuable information for some studies, it limits the user to the properties that have already been fitted or tabulated, like disk mass, disk size etc. Publicly available raw data of the disks after the stellar fly-by is still missing as of today. The database DESTINY provides this information so that the user can determine any disk property that may be of interest. We note that the data provided only accounts for effects after a fly-by and possible long-term effects like viscous spreading or planet-planet scattering are not included here.

In \cref{sec:simulations} we present a brief overview of the numerical method that has been employed, the model assumptions and limitations. Section \ref{sec:database} deals with the practicalities of using this database, like the parameter space covered (\cref{sec:parameters}), its structure (\cref{sec:structure}), downloading options (\cref{sec:download}), visualization tool (\cref{sec:graphics}), and advantages (\cref{sec:advantages}). Lastly, the possible applications of DESTINY to determine the effect of stellar encounters on disks as well as planetary systems are discussed in \cref{sec:applications}. 

\subsection{Model assumptions and numerical method}
\label{sec:simulations}
We performed numerical simulations using three-body interactions by only considering gravitational forces between a star surrounded by  a thin disk \citep{Pringle1981}, which we represent by 10\,000 mass-less tracer particles on Keplerian orbits and a second star perturbing the system via a fly-by. It has been shown in a number of studies that this resolution is sufficient for investigations of the global properties of disks \citep{Kobayashi2001,Pfalzner2003}.

The position and velocity of the particles are completely determined by using the orbital plane radius and the orbital phase (true anomaly). The trajectories of the particles during and after the stellar encounter were integrated with the Runge-Kutta Cash-Karp scheme; the maximum allowed error between the 4th and 5th integration step was $10^{-7}$. This integrator suffices because it allows for a statistical accuracy better than the typical 2-3\% error that arises from the sampling of the disc. We considered an inner hole of 1~au to avoid small time steps and to account for matter accreted onto the host star. 

These simulations can either be applied to a star surrounded by a disk or a planetary system. In the latter case the test particles are representative for all possible locations of planets. 

DESTINY is however applicable only to cases in which self-gravity and viscosity play a minor role and therefore can be neglected. This is automatically valid for debris disks, but one has to be careful with applications to protoplanetary disks and planetary systems that might become unstable at a later time.
 
Self-gravity can be neglected if the disk mass $m_{\mathrm{disk}}$ is small in comparison to the stellar masses $M_{*}$ involved in the fly-by which is valid for most observed disks \citep[see][]{Andrews2013}. Effects of viscous forces can be neglected for a star surrounded by planetary systems and it is often also applicable to disks because the encounter time is short compared to the viscous timescale. Equally, any property that mainly concerns the outer areas of the disk can be well described by the results given in the database. By contrast, results like fly-by induced accretion can only be regarded as first estimates. 

In the case of a low-mass, non-viscous disk and planetary systems, it suffices to study only three-body interactions by considering the gravitational forces between the two stars and each disk particle \citep{Hall1996, Kobayashi2001, Pfalzner2003, PfalznerVogel2005, Breslau2014, Musielak2014}. The simulations also only account for the effects immediately after the fly-by. Hence long-term effects like viscous spreading or planet-planet scattering can have additional effects on disk which are not included in the database provided here. 

The database represents encounter scenarios where only one of the stars is surrounded by a disk. In many cases the results from star-disk encounters can be generalized to disk-disk encounters by simply adding both components \citep{PfalznerUmbreit2005}. The captured mass is usually deposited in the inner disk areas and as such does not influence the final disk size \citep{PfalznerUmbreit2005}. The general trend for the effects on disk masses and accretion is similar. However, for very close fly-bys the actual values can increase by up to 10\%.

We note that this is purely a technical paper on how to use the database. Further details of the numerical method and a discussion on the applicability of these approximations can be found in \citep{Breslau2014, Bhandare2016}. 

\subsection{DESTINY }
\label{sec:database}

DESTINY \citep{DESTINY} provides access to the raw datasets which can be downloaded to perform one's own diagnostics. A readme file which describes the structure and contents of the datasets is provided. The website includes two short videos showing the typical fly-by dynamics. Furthermore, the website contains details of the scanned parameter space in a tabular format and allows the user to freely choose various parameters and view the effects of different types of encounters via a graphical interface. 

\begin{figure*}[tp]
\centering
Encounter scenarios
\includegraphics[width=\textwidth]{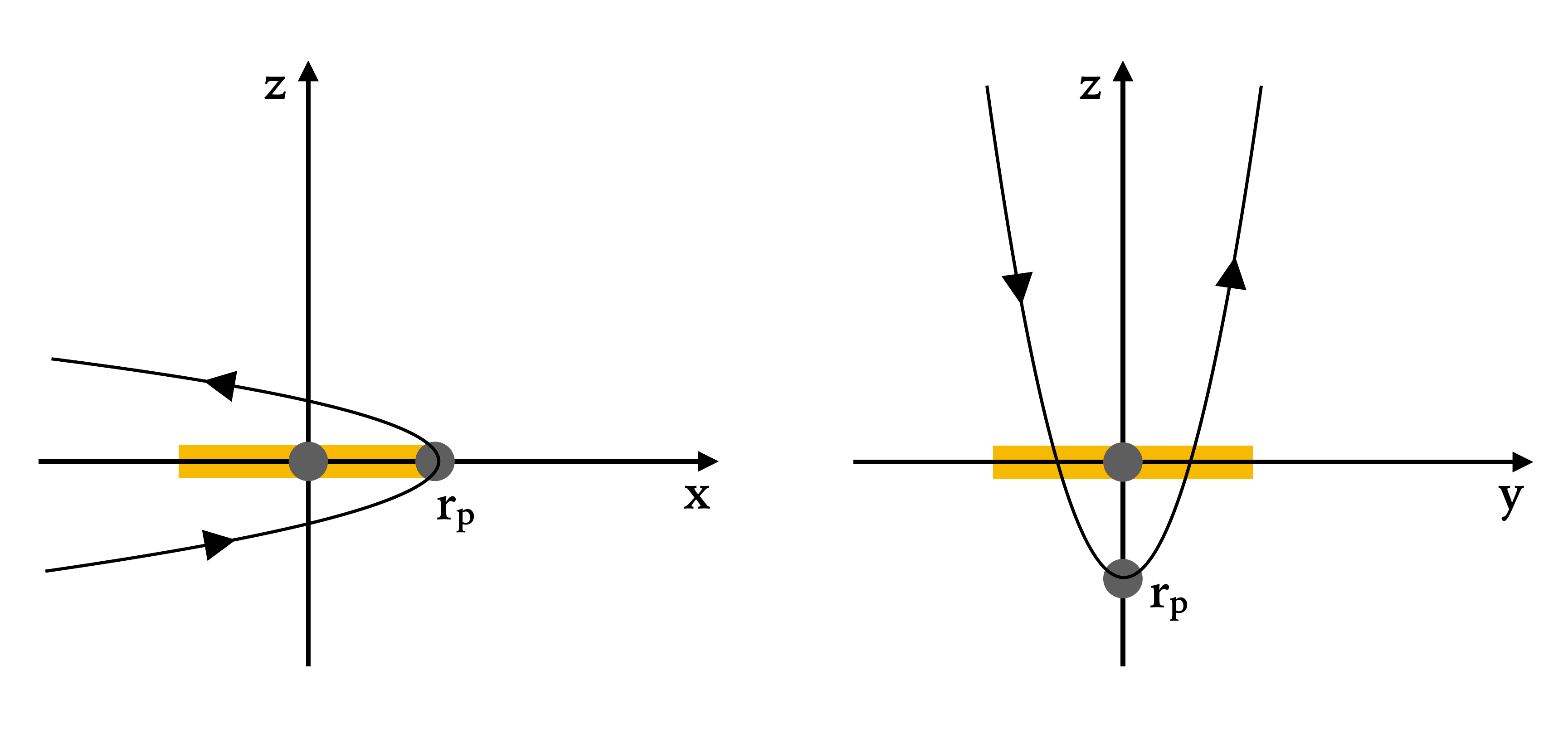}	
\caption{Encounter orbit with periastron $r_{\mathrm{p}}$ in the disk plane  \mbox{$\omega = 0^{\circ}$} (left) and below the disk plane $\omega = 90^{\circ}$ (right).}
\label{fig:encounter}
\end{figure*}

\subsubsection{Parameter space}
\label{sec:parameters}

Fly-by simulations as described in \cref{sec:simulations} were performed for ratios of perturber mass to host mass, $m_{\mathrm{12}}$ = $M_{\mathrm{2}}/M_{\mathrm{1}}$ in the range 0.3 -- 50. This corresponds to the range relevant for young dense clusters like the Orion nebula cluster (ONC) \citep{Weidner2010}. For mass ratios smaller than 0.3 the effects on disk sizes is smaller than the typical error ($\sim$ 2\%) in the simulations. For the low-mass perturbers one has to be aware that the disk mass is not significantly smaller than the mass of the perturbing star. In such cases additional effects due to pressure, viscous forces and self gravity can become important which have been neglected in these simulations.

In the case where self-gravity and viscosity can be neglected, the problem scales with the initial disk size, periastron distance and the mass ratio. This means the results presented here are applicable to any stellar mass and initial disk size by applying the following scaling laws 
\begin{align}
r_{\mathrm{final}} = 
\begin{cases}
0.28 \cdot {r_{\mathrm{peri}} \cdot {{m_{\mathrm{12}}}^{-0.32}}}, \hspace{2em}  & \text{for}  ~r_{\mathrm{final}} \leq r_{\mathrm{init}} \\
r_{\mathrm{init}}, & \text{otherwise},
\end{cases}
\label{eq:discsize_Breslau} 
\end{align} 
as presented in \citep{Breslau2014}. The data provided herein refers to the case of 1 $M_\odot$ star surrounded by an initial 100 au disk.

Periastron distances in the range $r_{\mathrm{peri}}$ = 30 -- 1000 au are studied to cover the parameter space from fly-bys that almost completely destroy the disk to those having a negligible effect on the disk size. We define disks to be completely destroyed when less than 5$\%$ of the  original disk mass remains bound whereas the effect of the fly-by is considered to be negligible if its effect on the disk size is less than 2\%.
 
Considering the disk to be in the xy plane, in principle the perturber orbit can be inclined in two ways, either along the x-axis wherein the periastron always lies in the disk plane or with respect to the xz plane wherein the periastron lies outside the disk plane. Thus the orbit of the perturbing star can be rotated in the disk plane, resulting in different angles between the periastron and the ascending node (here on the x-axis because the longitude of the ascending node is zero). Hence, we consider the effects of change in the argument of periapsis ($\omega$) as illustrated in Fig. \ref{fig:encounter}.

In addition to the angle of periastron, we also vary the inclination of the perturber orbit in the range \mbox{$0^{\circ} - {180^\circ}$} in steps of $10^{\circ}$ for each of the three cases of $\omega = 0^{\circ}$, $\omega = 45^{\circ}$, and $\omega = 90^{\circ}$ that we investigate. By doing so the entire parameter space to study both coplanar prograde ($i = 0^{\circ}$) $\&$ retrograde ($i = 180^{\circ}$) and also non-coplanar prograde ({$0^{\circ} < i < 90^{\circ}$}) $\&$ retrograde ({$90^{\circ} < i < 180^{\circ}$}) is covered. In addition, one can also study the effects due to an encounter with a perturber on an orthogonal ($i = 90^{\circ}$) orbit. This is an interesting case, since for encounters with $r_{\mathrm{peri}} < r_{\mathrm{init}}$ the perturber passes right through the disk without having interacted much with the disk material before and after it crosses the disk. We thus cover a wider range of orbital inclinations in comparison to previous studies \citep{Kobayashi2001, Kobayashi2005, Breslau2014}. The parameter space scanned in this study is listed in Table \ref{tab:setupparams} and can also be found on the website \citep{DESTINY}.

\begin{table}[t]
\centering
\caption{{Parameter space of the modelled fly-bys. Listed below is the mass ratio (perturber mass / host mass), periastron distances in au and the orbital inclination and orientation (angle of periastron) in degrees.}}
\resizebox{\textwidth}{!}{
\begin{tabular}[t]{l|l} \hline
Parameter & Simulated values  \TBstrut \\
\hline \hline
Mass ratio (perturber mass / host mass) & 0.3, 0.5, 1.0, 2.0, 5.0, 10.0, 20.0, 50.0  \Tstrut \\  
Angle of periastron [degrees] & 0, 45, 90  \\
Orbital inclination [degrees]  & 0, 10, 20, 30, 40, 50, 60, 70, 80, 90, 100, 110, 120, 130, 140, 150, 160, 170, 180 \\
Periastron distance [au] & 30, 50, 70, 100, 120, 150, 200, 250, 300, 500, 700, 1000 \Bstrut \\ \hline
\end{tabular}
}
\label{tab:setupparams}
\end{table}

For values other than those given in the database, an interpolation can be used to obtain the desired values. In almost all cases, linear interpolation will give reliable results. Only for orbital inclinations between 120 and 160, combined with short periastron distances there might be a problem, because there is a local minimum there. In this range, a quadratic fit might be more suitable.

In this database only fly-bys on parabolic orbits ($e_p$ = 1) have been included so far. (Near) parabolic encounters dominate for typical clusters in the solar neighborhood \citep{Olczak2006, Vincke2016}. The situation is different in very dense clusters like Westerlund 2 and Arches, where hyperbolic encounters are much more common \citep{Olczak2012,Vincke2018}. However, since disks are less affected by hyperbolic encounters in comparison to the parabolic ones, the effect of parabolic encounters can be regarded as an upper limit. The database will be extended to include fly-bys on hyperbolic orbits in the near future.

\subsubsection{Structure}
\label{sec:structure}
The datasets contain data in the hdf5\footnote{https://portal.hdfgroup.org/display/HDF5/HDF5} file format for the entire parameter space described in \cref{sec:parameters}.
HDF5 is a data model, library, and file format for storing and managing data, which is designed for flexible and efficient I/O and for high volume and complex data. 

\begin{figure*}[tp]
\centering
Database structure
\includegraphics[width=0.8\textwidth]{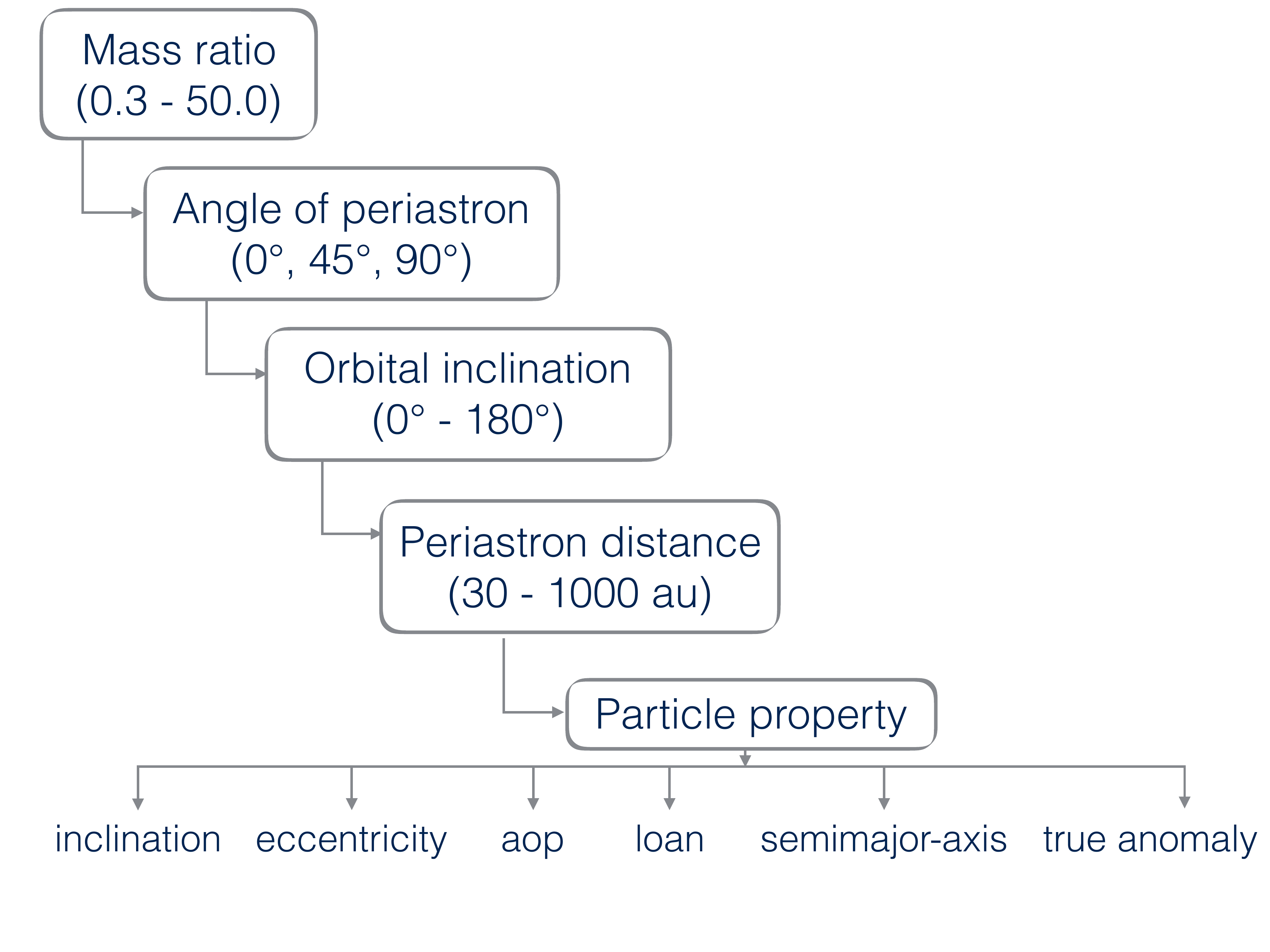}	
\caption{Shown here is the tree describing the structure of the datasets stored in hdf5 file format. The mass ratio is the main group and the rest are sub-groups. }
\label{fig:structure}
\end{figure*}

The data is structured in a hierarchical tree as illustrated in Fig. \ref{fig:structure}. The mass ratio is chosen to be the main group. Every main group has three sub-groups which contain data for three different angle of periastra. Each of these three sub-groups again are divided into 19 different sub-groups for each orbital inclination. Each of the 19 different sub-groups then contain 12 different sub-groups for different periastron distances.

\subsubsection{Download and usage of the raw data}
\label{sec:download}
The raw data can be found by clicking the button ``Datasets" available on the website. Apart from a readme file, eight different hdf5 files can be easily downloaded. The datasets named as ``dataset\_massratio.hdf5" contain data for the entire parameter space (angle of periastron, orbital inclination and periastron distance) for a particular mass ratio. For example the file  ``dataset\_m0000.3000.hdf5" contains particle properties for an encounter scenario where the mass ratio is 0.3 i.e. the perturber mass is 0.3 times that of the host star for each of the possible cases with different angle of periastra, orbital inclinations and periastron distances. Thus one can either download the complete set for the different mass ratios or choose a specific case that one is interested in.

For different fly-by scenarios the below listed properties of the particles can be accessed in each of the datasets. The readme file contains an example to view or save any of these particle properties, namely, 
\begin{itemize}
\item x, y, z positions: x\_position, y\_position, z\_position
\item inclination: particle\_inclination
\item eccentricty: particle\_eccentricity
\item semimajor axis: particle\_semimajoraxis
\item longitude of ascending node: particle\_loan
\item angle of periastron (or argument of periapsis): particle\_aop
\item true anomaly: particle\_true\_anomaly
\end{itemize} 

where the x, y, z positions are the final time averaged positions.

Currently only the data for the particles that are still bound to the host star after the fly-by are available in the datasets. However an outlook is to also include the unbound particles as well as those captured by the perturber. 

\subsubsection{Graphics options}
\label{sec:graphics}

Apart from the access to the raw data, the user is also provided with an option to visualize the outcome of a fly-by via the graphical interface. In this case no actual data download is necessary, but the user can choose any desired encounter scenario from a drop-down list for different mass ratios, angle of periastra (aop), orbital inclinations, periastron distances, and particle properties. A face-on and edge-on view of the fly-by effects is displayed showing the different particles (in an initially 100 au disk) that are finally bound to the host star, color coded with the selected particle property. 

\begin{figure}[tp]
\centering
Particle properties \\
\includegraphics[width=0.48\textwidth]{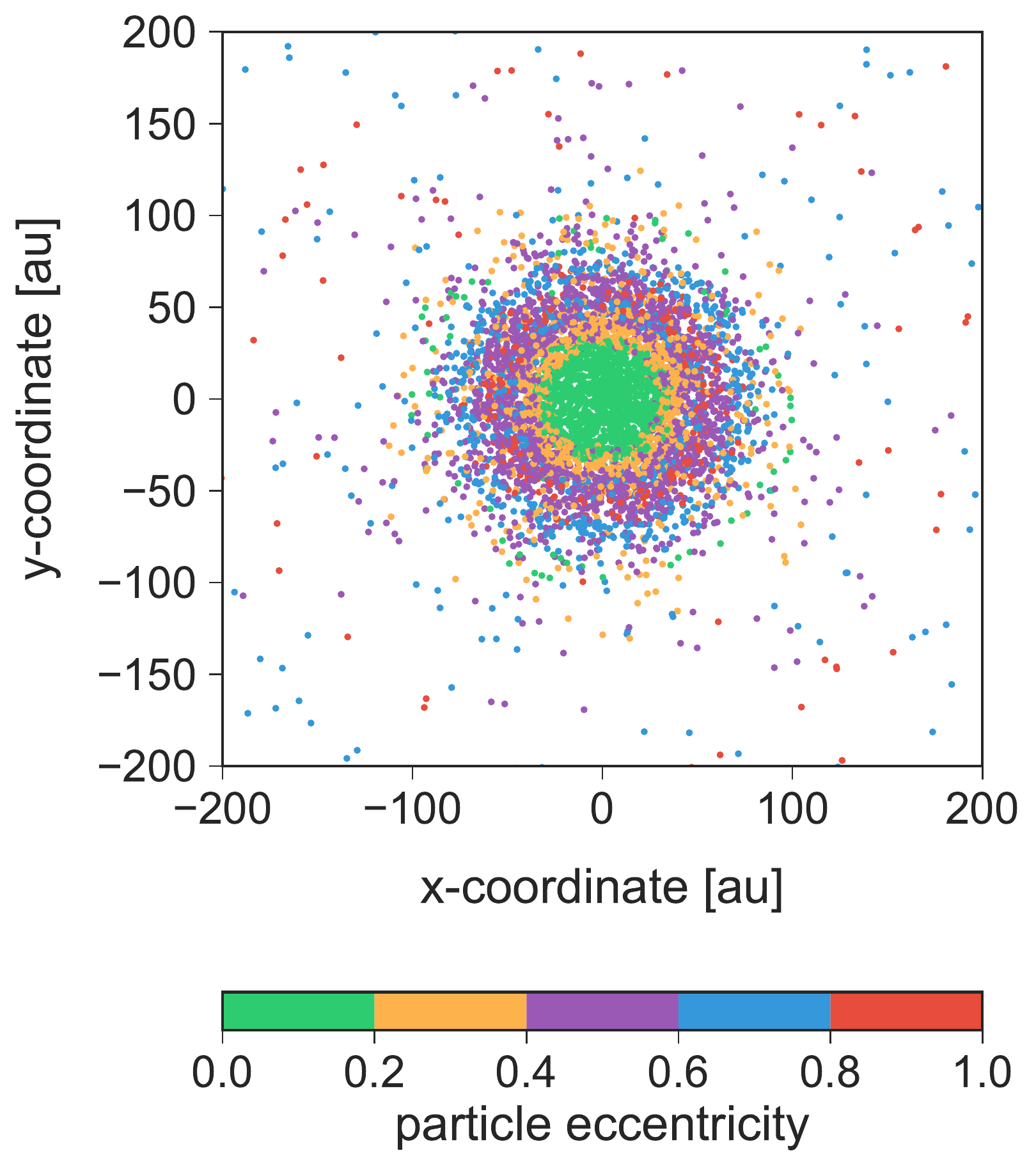}	
\includegraphics[width=0.48\textwidth]{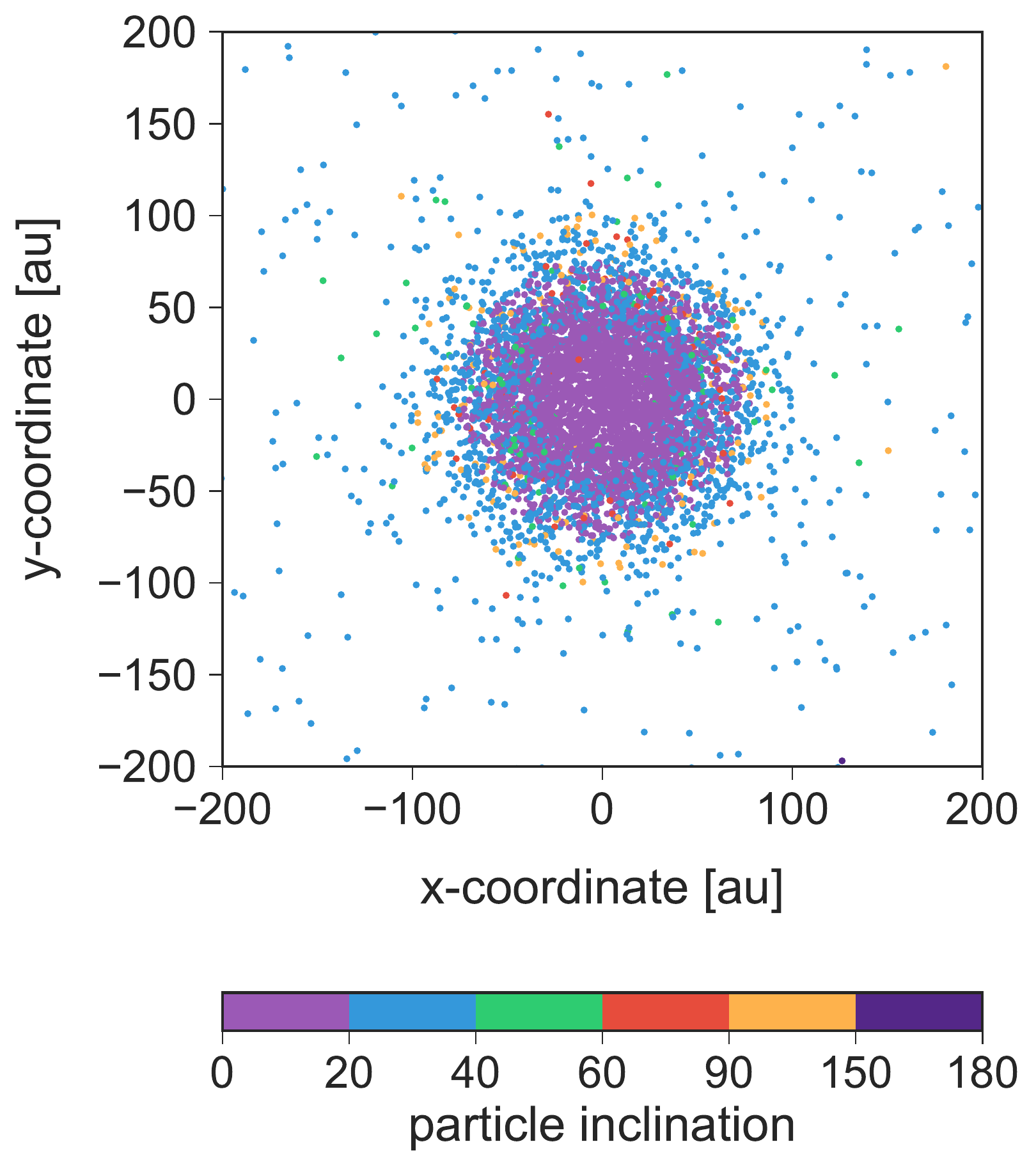}	
\caption{Face-on disk plots showing particle eccentricity (top) and particle inclination (bottom) at the final time step after an encounter at periastron distance of 100 au by a 1 $M_{\odot}$ perturber at orbital inclination of $60^{\circ}$ and angle of peristron of $0^{\circ}$.}
\label{fig:diskplots}
\end{figure}

Figure \ref{fig:diskplots} shows examples of face-on disks indicating the effects on the particle eccentricity and inclination after an encounter at a periastron distance of 100 au by a 1 $M_{\odot}$ perturber at orbital inclination of $60^{\circ}$ and angle of peristron of $0^{\circ}$. As seen in the figure, most of the inner disk particles remain coplanar and on circular orbits in comparison to the outer disk particles which are scattered on eccentric and inclined orbits. The graphical tool allows the user to quickly browse through the extensive dataset and visualize the different effects, thus making it easier to select interesting cases and then download the desired dataset to perform one's own diagnostics. 

\subsubsection{Advantages of the database}
\label{sec:advantages}
The work presented herein is an outcome of a collective effort of continuously increasing the accuracy and expanding parameter space over the last 15 years \citep{PfalznerVogel2005,Olczak2006,Steinhausen2012,Breslau2014,Bhandare2016,Vincke2018}. Compared to earlier works \citep{PfalznerVogel2005,Olczak2006} the data used for DESTINY has several advantages, namely,
\begin{itemize}
    \item higher accuracy,
    \item larger parameter space, 
    \item better resolution in the outer disk areas,
    \item free choice of initial disk density distribution,
    \item more flexibility for the user.
\end{itemize}

The higher accuracy is achieved mainly by simulating a larger time span before and after the periastron passage. The simulation starts and ends when it holds for all particles bound to the host that the force of the perturber on the particles is 0.1$\%$ less than that of the host star. As an example, the total simulation time for an equal-mass case then corresponds to around 40 orbits for the outermost particles and more than 50 orbits for the inner particles.

Breslau \citep{Breslau2014} studied the effect of coplanar encounters on the disk size for a wide range of mass ratios and periastron distances using three-body interactions. In our previous work \citep{Bhandare2016}, we extended this parameter space to also investigate the effects of inclined and retrograde parabolic encounters at different orientations. Thus the effect of a stellar encounters due to various properties such as the periastron distance, mass ratio between the perturber and the host star, the relative inclination and angle of periastron of the perturber orbit can be obtained.

In most fly-by simulations the initial disk-mass distribution is simulated by assigning each pseudo-particle the same mass \citep{Boffin1998,Pfalzner2004,Olczak2006,Pfalzner2007,ovelar2012}. In contrast, here a fixed test-particle distribution is used initially and the particles are assigned a mass according to the desired density distribution in the disk during post processing \citep{Breslau2014, Bhandare2016}. This has two advantages, first, such a flexible numerical scheme allows using one suite of simulations for any initial disk mass 
distribution and second, this method helps achieve a higher resolution particularly in the outer disk/planetary system areas. This is important for most applications as these outer disk/planetary system regions are affected by fly-bys the most.

Thus far only the information for the disk mass, angular momentum change, disk size etc. after a stellar fly-by has been provided either in tabulated format or in terms of a fit formula as a function of the fly-by parameters. DESTINY not only provides the user with this information but also allows access to the raw data for each individual test particle giving the freedom to choose the properties that one wants to investigate in this context. This means that the user can not only obtain pre-defined properties, but also derive independently the features they are interested in. In addition, the graphical tool allows the user to visualize various properties and understand the fate of disks after a stellar fly-by. All of the raw data from this extensive parameter study is publicly available online \citep{DESTINY}. 

\subsection{Applications}
\label{sec:applications}

One application of this database is the study of the effect of fly-bys on protoplanetary disks surrounding young stars similar to \citep{Thies2005,Kobayashi2005,Jilkova2015,Dai2015,Xiang2016,Mustill2016}. A recent example is the application to the solar system itself \citep{Pfalzner2018}. 

The database can not only be applied to protoplanetary disks but also to planetary systems. The planetary system around Pr 0211 in the M44 cluster \citep{Pfalzner2018a} is an example of one such study. So far, mostly the effect on planetary systems has been modelled by performing cluster simulations where each of the stars is surrounded by one or more planets \citep{Zheng2015,Dotti2018,Fujii2019}. The problem is that these simulations are computationally expensive and hence are restricted to specific planetary systems. Here, the parameter space covered by test particles is interpreted as potential locations of planets. The advantage compared to directly simulating the effect on specific planetary systems is that many planetary system configurations can be investigated simultaneously. This improves the statistical significance considerably. However, the prize to pay is that there is no knowledge about the long-term behaviour of the planetary system after the fly-by.

A third application is to investigate the effects in debris disks. Although the close fly-by frequency is very low for these systems, this is outbalanced by the fact that they can exist for hundreds of Myrs, possibly Gyrs. Despite their low frequency, close fly-bys can occur during such an extended time span. The effect of fly-bys on debris disks has been rarely studied so far, but could potentially provide an explanation for the observed asymmetries in debris disks.  

\subsection{Summary}
\label{sec:Summary}
In summary, the database DESTINY \citep{DESTINY} provides raw data and a graphical tool to investigate the effects of stellar encounters on disks. Such fly-bys are relatively common in the early phases of star clusters and associations. DESTINY covers a wide parameter range necessary for applications to disks as well as planetary systems in such different environments. The details of the contents, structure and potential usage of the database are discussed. Additionally, the assumptions made when acquiring the data and therefore also the limitations of the applications are clarified. 

\section{ABBREVIATIONS}
DESTINY: Database for the Effects of STellar encounters on dIsks and plaNetary sYstems.

\section{DECLARATIONS}

\subsection{Availability of data and material}
All of the data from this extensive parameter study is now publicly available online \citep{DESTINY}.

\subsection{Competing interests}
The authors declare that they have no competing interests.

\subsection{Funding}
Not applicable.

\subsection{Authors' contributions}
Both authors have equally contributed to the simulations performed to obtain the data available in this database and in writing the manuscript. All authors have read and approved the final manuscript.

\subsection{Acknowledgements}
We would like to thank Mikhail Kovalev for his help with setting up the webpage and the IT at MPIfR and Annika Hagemann for their help with hosting the webpage. This work would not have been possible without all the useful discussions with Andreas Breslau and Kirsten Vincke during this project. 


\let\jnlstyle=\rmfamily
\def\refjnl#1{{\jnlstyle#1}}%
\def\araa{ARA\&A}                      
\def\aj{AJ}                            
\def\apj{ApJ}                          
\def\apjs{ApJS}                        
\def\nat{Nature}                       
\newcommand\apjl{\refjnl{ApJ}}         
\newcommand\icarus{\refjnl{Icarus}}    
\newcommand\aap{\refjnl{A\&A}}         
\newcommand\aapr{\refjnl{A\&A~Rev.}}   
\newcommand\mnras{\refjnl{MNRAS}}      
\newcommand\eas{\refjnl{EAS}}          
\newcommand\apss{\refjnl{Ap\&SS}}      

\bibliographystyle{bmc-mathphys} 

\bibliography{Bibliography}      


\begin{thebibliography}{68}
\ifx \bisbn   \undefined \def \bisbn  #1{ISBN #1}\fi
\ifx \binits  \undefined \def \binits#1{#1}\fi
\ifx \bauthor  \undefined \def \bauthor#1{#1}\fi
\ifx \batitle  \undefined \def \batitle#1{#1}\fi
\ifx \bjtitle  \undefined \def \bjtitle#1{#1}\fi
\ifx \bvolume  \undefined \def \bvolume#1{\textbf{#1}}\fi
\ifx \byear  \undefined \def \byear#1{#1}\fi
\ifx \bissue  \undefined \def \bissue#1{#1}\fi
\ifx \bfpage  \undefined \def \bfpage#1{#1}\fi
\ifx \blpage  \undefined \def \blpage #1{#1}\fi
\ifx \burl  \undefined \def \burl#1{\textsf{#1}}\fi
\ifx \doiurl  \undefined \def \doiurl#1{\textsf{#1}}\fi
\ifx \betal  \undefined \def \betal{\textit{et al.}}\fi
\ifx \binstitute  \undefined \def \binstitute#1{#1}\fi
\ifx \binstitutionaled  \undefined \def \binstitutionaled#1{#1}\fi
\ifx \bctitle  \undefined \def \bctitle#1{#1}\fi
\ifx \beditor  \undefined \def \beditor#1{#1}\fi
\ifx \bpublisher  \undefined \def \bpublisher#1{#1}\fi
\ifx \bbtitle  \undefined \def \bbtitle#1{#1}\fi
\ifx \bedition  \undefined \def \bedition#1{#1}\fi
\ifx \bseriesno  \undefined \def \bseriesno#1{#1}\fi
\ifx \blocation  \undefined \def \blocation#1{#1}\fi
\ifx \bsertitle  \undefined \def \bsertitle#1{#1}\fi
\ifx \bsnm \undefined \def \bsnm#1{#1}\fi
\ifx \bsuffix \undefined \def \bsuffix#1{#1}\fi
\ifx \bparticle \undefined \def \bparticle#1{#1}\fi
\ifx \barticle \undefined \def \barticle#1{#1}\fi
\ifx \bconfdate \undefined \def \bconfdate #1{#1}\fi
\ifx \botherref \undefined \def \botherref #1{#1}\fi
\ifx \url \undefined \def \url#1{\textsf{#1}}\fi
\ifx \bchapter \undefined \def \bchapter#1{#1}\fi
\ifx \bbook \undefined \def \bbook#1{#1}\fi
\ifx \bcomment \undefined \def \bcomment#1{#1}\fi
\ifx \oauthor \undefined \def \oauthor#1{#1}\fi
\ifx \citeauthoryear \undefined \def \citeauthoryear#1{#1}\fi
\ifx \endbibitem  \undefined \def \endbibitem {}\fi
\ifx \bconflocation  \undefined \def \bconflocation#1{#1}\fi
\ifx \arxivurl  \undefined \def \arxivurl#1{\textsf{#1}}\fi
\csname PreBibitemsHook\endcsname

\bibitem{Moor2013}
\begin{barticle}
\bauthor{\bsnm{{Mo{\'o}r}}, \binits{A.}},
\bauthor{\bsnm{{Juh{\'a}sz}}, \binits{A.}},
\bauthor{\bsnm{{K{\'o}sp{\'a}l}}, \binits{{\'A}.}},
\bauthor{\bsnm{{{\'A}brah{\'a}m}}, \binits{P.}},
\bauthor{\bsnm{{Apai}}, \binits{D.}},
\bauthor{\bsnm{{Csengeri}}, \binits{T.}},
\bauthor{\bsnm{{Grady}}, \binits{C.}},
\bauthor{\bsnm{{Henning}}, \binits{T.}},
\bauthor{\bsnm{{Hughes}}, \binits{A.M.}},
\bauthor{\bsnm{{Kiss}}, \binits{C.}},
\bauthor{\bsnm{{Pascucci}}, \binits{I.}},
\bauthor{\bsnm{{Schmalzl}}, \binits{M.}},
\bauthor{\bsnm{{Gab{\'a}nyi}}, \binits{K.}}:
\batitle{{ALMA Continuum Observations of a 30 Myr Old Gaseous Debris Disk
  around HD 21997}}.
\bjtitle{\apjl}
\bvolume{777},
\bfpage{25}
(\byear{2013}).
doi:\doiurl{10.1088/2041-8205}.
\arxivurl{1310.5069}
\end{barticle}
\endbibitem

\bibitem{Mann2014}
\begin{barticle}
\bauthor{\bsnm{{Mann}}, \binits{R.K.}},
\bauthor{\bsnm{{Di Francesco}}, \binits{J.}},
\bauthor{\bsnm{{Johnstone}}, \binits{D.}},
\bauthor{\bsnm{{Andrews}}, \binits{S.M.}},
\bauthor{\bsnm{{Williams}}, \binits{J.P.}},
\bauthor{\bsnm{{Bally}}, \binits{J.}},
\bauthor{\bsnm{{Ricci}}, \binits{L.}},
\bauthor{\bsnm{{Hughes}}, \binits{A.M.}},
\bauthor{\bsnm{{Matthews}}, \binits{B.C.}}:
\batitle{{ALMA Observations of the Orion Proplyds}}.
\bjtitle{\apj}
\bvolume{784},
\bfpage{82}
(\byear{2014}).
doi:\doiurl{10.1088/0004-637X}.
\arxivurl{1403.2026}
\end{barticle}
\endbibitem

\bibitem{Bally2015}
\begin{barticle}
\bauthor{\bsnm{{Bally}}, \binits{J.}},
\bauthor{\bsnm{{Mann}}, \binits{R.K.}},
\bauthor{\bsnm{{Eisner}}, \binits{J.}},
\bauthor{\bsnm{{Andrews}}, \binits{S.M.}},
\bauthor{\bsnm{{Di Francesco}}, \binits{J.}},
\bauthor{\bsnm{{Hughes}}, \binits{M.}},
\bauthor{\bsnm{{Johnstone}}, \binits{D.}},
\bauthor{\bsnm{{Matthews}}, \binits{B.}},
\bauthor{\bsnm{{Ricci}}, \binits{L.}},
\bauthor{\bsnm{{Williams}}, \binits{J.P.}}:
\batitle{{ALMA Observations of the Largest Proto-Planetary Disk in the Orion
  Nebula, 114-426: A CO Silhouette}}.
\bjtitle{\apj}
\bvolume{808},
\bfpage{69}
(\byear{2015}).
doi:\doiurl{10.1088/0004-637X/808/1/69}.
\arxivurl{1506.03391}
\end{barticle}
\endbibitem

\bibitem{Tobin2015}
\begin{bchapter}
\bauthor{\bsnm{{Tobin}}, \binits{J.J.}},
\bauthor{\bsnm{{Looney}}, \binits{L.W.}},
\bauthor{\bsnm{{Li}}, \binits{Z.-Y.}},
\bauthor{\bsnm{{Chandler}}, \binits{C.J.}},
\bauthor{\bsnm{{Dunham}}, \binits{M.M.}},
\bauthor{\bsnm{{Segura-Cox}}, \binits{D.}},
\bauthor{\bsnm{{Cox}}, \binits{E.G.}},
\bauthor{\bsnm{{Harris}}, \binits{R.J.}},
\bauthor{\bsnm{{Melis}}, \binits{C.}},
\bauthor{\bsnm{{Sadavoy}}, \binits{S.I.}},
\bauthor{\bsnm{{P{\'e}rez}}, \binits{L.}},
\bauthor{\bsnm{{Kratter}}, \binits{K.}}:
\bctitle{{Revolutionizing our View of Protostellar Multiplicity and Disks: The
  VLA Nascent Disk and Multiplicity (VANDAM) Survey of the Perseus Molecular
  Cloud}}.
In: \bbtitle{EAS Publications Series}.
\bsertitle{EAS Publications Series},
vol. \bseriesno{75-76},
pp. \bfpage{273}--\blpage{276}
(\byear{2015}).
doi:\doiurl{10.1051/eas/1575054}.
\arxivurl{1607.01425}
\end{bchapter}
\endbibitem

\bibitem{Andrews2018}
\begin{barticle}
\bauthor{\bsnm{{Andrews}}, \binits{S.M.}},
\bauthor{\bsnm{{Huang}}, \binits{J.}},
\bauthor{\bsnm{{P{\'e}rez}}, \binits{L.M.}},
\bauthor{\bsnm{{Isella}}, \binits{A.}},
\bauthor{\bsnm{{Dullemond}}, \binits{C.P.}},
\bauthor{\bsnm{{Kurtovic}}, \binits{N.T.}},
\bauthor{\bsnm{{Guzm{\'a}n}}, \binits{V.V.}},
\bauthor{\bsnm{{Carpenter}}, \binits{J.M.}},
\bauthor{\bsnm{{Wilner}}, \binits{D.J.}},
\bauthor{\bsnm{{Zhang}}, \binits{S.}},
\bauthor{\bsnm{{Zhu}}, \binits{Z.}},
\bauthor{\bsnm{{Birnstiel}}, \binits{T.}},
\bauthor{\bsnm{{Bai}}, \binits{X.-N.}},
\bauthor{\bsnm{{Benisty}}, \binits{M.}},
\bauthor{\bsnm{{Hughes}}, \binits{A.M.}},
\bauthor{\bsnm{{{\"O}berg}}, \binits{K.I.}},
\bauthor{\bsnm{{Ricci}}, \binits{L.}}:
\batitle{{The Disk Substructures at High Angular Resolution Project (DSHARP).
  I. Motivation, Sample, Calibration, and Overview}}.
\bjtitle{\apj}
\bvolume{869},
\bfpage{41}
(\byear{2018}).
doi:\doiurl{10.3847/2041-8213/aaf741}.
\arxivurl{1812.04040}
\end{barticle}
\endbibitem

\bibitem{Bromley2011}
\begin{barticle}
\bauthor{\bsnm{{Bromley}}, \binits{B.C.}},
\bauthor{\bsnm{{Kenyon}}, \binits{S.J.}}:
\batitle{{A New Hybrid N-body-coagulation Code for the Formation of Gas Giant
  Planets}}.
\bjtitle{\apj}
\bvolume{731},
\bfpage{101}
(\byear{2011}).
doi:\doiurl{10.1088/0004-637X/731/2/101}.
\arxivurl{1012.0574}
\end{barticle}
\endbibitem

\bibitem{Baruteau2014}
\begin{bchapter}
\bauthor{\bsnm{{Baruteau}}, \binits{C.}},
\bauthor{\bsnm{{Crida}}, \binits{A.}},
\bauthor{\bsnm{{Paardekooper}}, \binits{S.-J.}},
\bauthor{\bsnm{{Masset}}, \binits{F.}},
\bauthor{\bsnm{{Guilet}}, \binits{J.}},
\bauthor{\bsnm{{Bitsch}}, \binits{B.}},
\bauthor{\bsnm{{Nelson}}, \binits{R.}},
\bauthor{\bsnm{{Kley}}, \binits{W.}},
\bauthor{\bsnm{{Papaloizou}}, \binits{J.}}:
\bctitle{{Planet-Disk Interactions and Early Evolution of Planetary Systems}}.
In: \beditor{\bsnm{{Beuther}}, \binits{H.}},
\beditor{\bsnm{{Klessen}}, \binits{R.S.}},
\beditor{\bsnm{{Dullemond}}, \binits{C.P.}},
\beditor{\bsnm{{Henning}}, \binits{T.}} (eds.)
\bbtitle{Protostars and Planets VI},
p. \bfpage{667}
(\byear{2014}).
doi:\doiurl{10.2458/azu_uapress_9780816531240-ch029}.
\arxivurl{1312.4293}
\end{bchapter}
\endbibitem

\bibitem{Bitsch2015}
\begin{barticle}
\bauthor{\bsnm{{Bitsch}}, \binits{B.}},
\bauthor{\bsnm{{Johansen}}, \binits{A.}},
\bauthor{\bsnm{{Lambrechts}}, \binits{M.}},
\bauthor{\bsnm{{Morbidelli}}, \binits{A.}}:
\batitle{{The structure of protoplanetary discs around evolving young stars}}.
\bjtitle{\aap}
\bvolume{575},
\bfpage{28}
(\byear{2015}).
doi:\doiurl{10.1051/0004-6361/201424964}.
\arxivurl{1411.3255}
\end{barticle}
\endbibitem

\bibitem{Bromley2015}
\begin{barticle}
\bauthor{\bsnm{{Bromley}}, \binits{B.C.}},
\bauthor{\bsnm{{Kenyon}}, \binits{S.J.}}:
\batitle{{Planet Formation around Binary Stars: Tatooine Made Easy}}.
\bjtitle{\apj}
\bvolume{806},
\bfpage{98}
(\byear{2015}).
doi:\doiurl{10.1088/0004-637X/806/1/98}.
\arxivurl{1503.03876}
\end{barticle}
\endbibitem

\bibitem{Clarke2000}
\begin{botherref}
\oauthor{\bsnm{{Clarke}}, \binits{C.J.}},
\oauthor{\bsnm{{Bonnell}}, \binits{I.A.}},
\oauthor{\bsnm{{Hillenbrand}}, \binits{L.A.}}:
{The Formation of Stellar Clusters}.
Protostars and Planets IV,
151
(2000).
\arxivurl{astro-ph/9903323}
\end{botherref}
\endbibitem

\bibitem{Lada2003}
\begin{barticle}
\bauthor{\bsnm{{Lada}}, \binits{C.J.}},
\bauthor{\bsnm{{Lada}}, \binits{E.A.}}:
\batitle{{Embedded Clusters in Molecular Clouds}}.
\bjtitle{\araa}
\bvolume{41},
\bfpage{57}--\blpage{115}
(\byear{2003}).
doi:\doiurl{10.1146/annurev.astro.41.011802.094844}.
\arxivurl{astro-ph/0301540}
\end{barticle}
\endbibitem

\bibitem{Porras2003}
\begin{barticle}
\bauthor{\bsnm{{Porras}}, \binits{A.}},
\bauthor{\bsnm{{Christopher}}, \binits{M.}},
\bauthor{\bsnm{{Allen}}, \binits{L.}},
\bauthor{\bsnm{{Di Francesco}}, \binits{J.}},
\bauthor{\bsnm{{Megeath}}, \binits{S.T.}},
\bauthor{\bsnm{{Myers}}, \binits{P.C.}}:
\batitle{{A Catalog of Young Stellar Groups and Clusters within 1 Kiloparsec of
  the Sun}}.
\bjtitle{\aj}
\bvolume{126},
\bfpage{1916}--\blpage{1924}
(\byear{2003}).
doi:\doiurl{10.1086/377623}.
\arxivurl{astro-ph/0307510}
\end{barticle}
\endbibitem

\bibitem{Hollenbach2000}
\begin{botherref}
\oauthor{\bsnm{{Hollenbach}}, \binits{D.J.}},
\oauthor{\bsnm{{Yorke}}, \binits{H.W.}},
\oauthor{\bsnm{{Johnstone}}, \binits{D.}}:
{Disk Dispersal around Young Stars}.
Protostars and Planets IV,
401
(2000)
\end{botherref}
\endbibitem

\bibitem{Williams2011}
\begin{barticle}
\bauthor{\bsnm{{Williams}}, \binits{J.P.}},
\bauthor{\bsnm{{Cieza}}, \binits{L.A.}}:
\batitle{{Protoplanetary Disks and Their Evolution}}.
\bjtitle{\araa}
\bvolume{49},
\bfpage{67}--\blpage{117}
(\byear{2011}).
doi:\doiurl{10.1146/annurev-astro-081710-102548}.
\arxivurl{1103.0556}
\end{barticle}
\endbibitem

\bibitem{Johnstone1998}
\begin{barticle}
\bauthor{\bsnm{{Johnstone}}, \binits{D.}},
\bauthor{\bsnm{{Hollenbach}}, \binits{D.}},
\bauthor{\bsnm{{Bally}}, \binits{J.}}:
\batitle{{Photoevaporation of Disks and Clumps by Nearby Massive Stars:
  Application to Disk Destruction in the Orion Nebula}}.
\bjtitle{\apj}
\bvolume{499},
\bfpage{758}--\blpage{776}
(\byear{1998})
\end{barticle}
\endbibitem

\bibitem{Adams2004}
\begin{barticle}
\bauthor{\bsnm{{Adams}}, \binits{F.C.}},
\bauthor{\bsnm{{Hollenbach}}, \binits{D.}},
\bauthor{\bsnm{{Laughlin}}, \binits{G.}},
\bauthor{\bsnm{{Gorti}}, \binits{U.}}:
\batitle{{Photoevaporation of Circumstellar Disks Due to External
  Far-Ultraviolet Radiation in Stellar Aggregates}}.
\bjtitle{\apj}
\bvolume{611},
\bfpage{360}--\blpage{379}
(\byear{2004}).
doi:\doiurl{10.1086/421989}.
\arxivurl{astro-ph/0404383}
\end{barticle}
\endbibitem

\bibitem{Font2004}
\begin{barticle}
\bauthor{\bsnm{{Font}}, \binits{A.S.}},
\bauthor{\bsnm{{McCarthy}}, \binits{I.G.}},
\bauthor{\bsnm{{Johnstone}}, \binits{D.}},
\bauthor{\bsnm{{Ballantyne}}, \binits{D.R.}}:
\batitle{{Photoevaporation of Circumstellar Disks around Young Stars}}.
\bjtitle{\apj}
\bvolume{607},
\bfpage{890}--\blpage{903}
(\byear{2004}).
doi:\doiurl{10.1086/383518}.
\arxivurl{astro-ph/0402241}
\end{barticle}
\endbibitem

\bibitem{Clarke2007}
\begin{barticle}
\bauthor{\bsnm{{Clarke}}, \binits{C.J.}}:
\batitle{{The photoevaporation of discs around young stars in massive
  clusters}}.
\bjtitle{\mnras}
\bvolume{376},
\bfpage{1350}--\blpage{1356}
(\byear{2007}).
doi:\doiurl{10.1111/j.1365-2966.2007.11547.x}.
\arxivurl{astro-ph/0702112}
\end{barticle}
\endbibitem

\bibitem{Dullemond2007}
\begin{botherref}
\oauthor{\bsnm{{Dullemond}}, \binits{C.P.}},
\oauthor{\bsnm{{Hollenbach}}, \binits{D.}},
\oauthor{\bsnm{{Kamp}}, \binits{I.}},
\oauthor{\bsnm{{D'Alessio}}, \binits{P.}}:
{Models of the Structure and Evolution of Protoplanetary Disks}.
Protostars and Planets V,
555--572
(2007).
\arxivurl{astro-ph/0602619}
\end{botherref}
\endbibitem

\bibitem{Gorti2009}
\begin{barticle}
\bauthor{\bsnm{{Gorti}}, \binits{U.}},
\bauthor{\bsnm{{Hollenbach}}, \binits{D.}}:
\batitle{{Photoevaporation of Circumstellar Disks By Far-Ultraviolet,
  Extreme-Ultraviolet and X-Ray Radiation from the Central Star}}.
\bjtitle{\apj}
\bvolume{690},
\bfpage{1539}--\blpage{1552}
(\byear{2009}).
doi:\doiurl{10.1088/0004-637X/690/2/1539}.
\arxivurl{0809.1494}
\end{barticle}
\endbibitem

\bibitem{Owen2010}
\begin{barticle}
\bauthor{\bsnm{{Owen}}, \binits{J.E.}},
\bauthor{\bsnm{{Ercolano}}, \binits{B.}},
\bauthor{\bsnm{{Clarke}}, \binits{C.J.}},
\bauthor{\bsnm{{Alexander}}, \binits{R.D.}}:
\batitle{{Radiation-hydrodynamic models of X-ray and EUV photoevaporating
  protoplanetary discs}}.
\bjtitle{\mnras}
\bvolume{401},
\bfpage{1415}--\blpage{1428}
(\byear{2010}).
doi:\doiurl{10.1111/j.1365-2966.2009.15771.x}.
\arxivurl{0909.4309}
\end{barticle}
\endbibitem

\bibitem{Owen2012}
\begin{barticle}
\bauthor{\bsnm{{Owen}}, \binits{J.E.}},
\bauthor{\bsnm{{Clarke}}, \binits{C.J.}},
\bauthor{\bsnm{{Ercolano}}, \binits{B.}}:
\batitle{{On the theory of disc photoevaporation}}.
\bjtitle{\mnras}
\bvolume{422},
\bfpage{1880}--\blpage{1901}
(\byear{2012}).
doi:\doiurl{10.1111/j.1365-2966.2011.20337.x}.
\arxivurl{1112.1087}
\end{barticle}
\endbibitem

\bibitem{Rosotti2015}
\begin{barticle}
\bauthor{\bsnm{{Rosotti}}, \binits{G.P.}},
\bauthor{\bsnm{{Ercolano}}, \binits{B.}},
\bauthor{\bsnm{{Owen}}, \binits{J.E.}}:
\batitle{{The long-term evolution of photoevaporating transition discs with
  giant planets}}.
\bjtitle{\mnras}
\bvolume{454},
\bfpage{2173}--\blpage{2182}
(\byear{2015}).
doi:\doiurl{10.1093/mnras/stv2102}.
\arxivurl{1509.04278}
\end{barticle}
\endbibitem

\bibitem{Wijnen2017}
\begin{barticle}
\bauthor{\bsnm{{Wijnen}}, \binits{T.P.G.}},
\bauthor{\bsnm{{Pols}}, \binits{O.R.}},
\bauthor{\bsnm{{Pelupessy}}, \binits{F.I.}},
\bauthor{\bsnm{{Portegies Zwart}}, \binits{S.}}:
\batitle{{Disc truncation in embedded star clusters: Dynamical encounters
  versus face-on accretion}}.
\bjtitle{\aap}
\bvolume{604},
\bfpage{91}
(\byear{2017}).
doi:\doiurl{10.1051/0004-6361/201731072}.
\arxivurl{1706.07048}
\end{barticle}
\endbibitem

\bibitem{Heller1995}
\begin{barticle}
\bauthor{\bsnm{{Heller}}, \binits{C.H.}}:
\batitle{{Encounters with Protostellar Disks. II. Disruption and Binary
  Formation}}.
\bjtitle{\apj}
\bvolume{455},
\bfpage{252}
(\byear{1995}).
doi:\doiurl{10.1086/176573}
\end{barticle}
\endbibitem

\bibitem{Hall1996}
\begin{barticle}
\bauthor{\bsnm{{Hall}}, \binits{S.M.}},
\bauthor{\bsnm{{Clarke}}, \binits{C.J.}},
\bauthor{\bsnm{{Pringle}}, \binits{J.E.}}:
\batitle{{Energetics of star-disc encounters in the non-linear regime}}.
\bjtitle{\mnras}
\bvolume{278},
\bfpage{303}--\blpage{320}
(\byear{1996}).
\arxivurl{astro-ph/9510153}
\end{barticle}
\endbibitem

\bibitem{Clarke1993}
\begin{barticle}
\bauthor{\bsnm{{Clarke}}, \binits{C.J.}},
\bauthor{\bsnm{{Pringle}}, \binits{J.E.}}:
\batitle{{Accretion disc response to a stellar fly-by}}.
\bjtitle{\mnras}
\bvolume{261},
\bfpage{190}--\blpage{202}
(\byear{1993})
\end{barticle}
\endbibitem

\bibitem{PfalznerVogel2005}
\begin{barticle}
\bauthor{\bsnm{{Pfalzner}}, \binits{S.}},
\bauthor{\bsnm{{Vogel}}, \binits{P.}},
\bauthor{\bsnm{{Scharw{\"a}chter}}, \binits{J.}},
\bauthor{\bsnm{{Olczak}}, \binits{C.}}:
\batitle{{Parameter study of star-disc encounters}}.
\bjtitle{\aap}
\bvolume{437},
\bfpage{967}--\blpage{976}
(\byear{2005}).
doi:\doiurl{10.1051/0004-6361:20042467}.
\arxivurl{astro-ph/0504288}
\end{barticle}
\endbibitem

\bibitem{Kobayashi2001}
\begin{barticle}
\bauthor{\bsnm{{Kobayashi}}, \binits{H.}},
\bauthor{\bsnm{{Ida}}, \binits{S.}}:
\batitle{{The Effects of a Stellar Encounter on a Planetesimal Disk}}.
\bjtitle{\icarus}
\bvolume{153},
\bfpage{416}--\blpage{429}
(\byear{2001}).
doi:\doiurl{10.1006/icar.2001.6700}.
\arxivurl{astro-ph/0107086}
\end{barticle}
\endbibitem

\bibitem{Kobayashi2005}
\begin{barticle}
\bauthor{\bsnm{{Kobayashi}}, \binits{H.}},
\bauthor{\bsnm{{Ida}}, \binits{S.}},
\bauthor{\bsnm{{Tanaka}}, \binits{H.}}:
\batitle{{The evidence of an early stellar encounter in Edgeworth Kuiper
  belt}}.
\bjtitle{\icarus}
\bvolume{177},
\bfpage{246}--\blpage{255}
(\byear{2005}).
doi:\doiurl{10.1016/j.icarus.2005.02.017}
\end{barticle}
\endbibitem

\bibitem{Breslau2014}
\begin{barticle}
\bauthor{\bsnm{{Breslau}}, \binits{A.}},
\bauthor{\bsnm{{Steinhausen}}, \binits{M.}},
\bauthor{\bsnm{{Vincke}}, \binits{K.}},
\bauthor{\bsnm{{Pfalzner}}, \binits{S.}}:
\batitle{{Sizes of protoplanetary discs after star-disc encounters}}.
\bjtitle{\aap}
\bvolume{565},
\bfpage{130}
(\byear{2014}).
doi:\doiurl{10.1051/0004-6361/201323043}.
\arxivurl{1403.8099}
\end{barticle}
\endbibitem

\bibitem{Jilkova2016}
\begin{barticle}
\bauthor{\bsnm{{J{\'{\i}}lkov{\'a}}}, \binits{L.}},
\bauthor{\bsnm{{Hamers}}, \binits{A.S.}},
\bauthor{\bsnm{{Hammer}}, \binits{M.}},
\bauthor{\bsnm{{Portegies Zwart}}, \binits{S.}}:
\batitle{{Mass transfer between debris discs during close stellar encounters}}.
\bjtitle{\mnras}
\bvolume{457},
\bfpage{4218}--\blpage{4235}
(\byear{2016}).
doi:\doiurl{10.1093/mnras/stw264}.
\arxivurl{1601.08171}
\end{barticle}
\endbibitem

\bibitem{Bhandare2016}
\begin{barticle}
\bauthor{\bsnm{{Bhandare}}, \binits{A.}},
\bauthor{\bsnm{{Breslau}}, \binits{A.}},
\bauthor{\bsnm{{Pfalzner}}, \binits{S.}}:
\batitle{{Effects of inclined star-disk encounter on protoplanetary disk
  size}}.
\bjtitle{\aap}
\bvolume{594},
\bfpage{53}
(\byear{2016}).
doi:\doiurl{10.1051/0004-6361/201628086}.
\arxivurl{1608.03239}
\end{barticle}
\endbibitem

\bibitem{Winter2018a}
\begin{barticle}
\bauthor{\bsnm{{Winter}}, \binits{A.J.}},
\bauthor{\bsnm{{Clarke}}, \binits{C.J.}},
\bauthor{\bsnm{{Rosotti}}, \binits{G.}},
\bauthor{\bsnm{{Booth}}, \binits{R.A.}}:
\batitle{{Protoplanetary disc response to distant tidal encounters in stellar
  clusters}}.
\bjtitle{\mnras}
\bvolume{475},
\bfpage{2314}--\blpage{2325}
(\byear{2018}).
doi:\doiurl{10.1093/mnras/sty012}.
\arxivurl{1801.03510}
\end{barticle}
\endbibitem

\bibitem{Winter2018b}
\begin{barticle}
\bauthor{\bsnm{{Winter}}, \binits{A.J.}},
\bauthor{\bsnm{{Clarke}}, \binits{C.J.}},
\bauthor{\bsnm{{Rosotti}}, \binits{G.}},
\bauthor{\bsnm{{Ih}}, \binits{J.}},
\bauthor{\bsnm{{Facchini}}, \binits{S.}},
\bauthor{\bsnm{{Haworth}}, \binits{T.J.}}:
\batitle{{Protoplanetary disc truncation mechanisms in stellar clusters:
  comparing external photoevaporation and tidal encounters}}.
\bjtitle{\mnras}
\bvolume{478},
\bfpage{2700}--\blpage{2722}
(\byear{2018}).
doi:\doiurl{10.1093/mnras/sty984}.
\arxivurl{1804.00013}
\end{barticle}
\endbibitem

\bibitem{Cuello2019}
\begin{barticle}
\bauthor{\bsnm{{Cuello}}, \binits{N.}},
\bauthor{\bsnm{{Dipierro}}, \binits{G.}},
\bauthor{\bsnm{{Mentiplay}}, \binits{D.}},
\bauthor{\bsnm{{Price}}, \binits{D.J.}},
\bauthor{\bsnm{{Pinte}}, \binits{C.}},
\bauthor{\bsnm{{Cuadra}}, \binits{J.}},
\bauthor{\bsnm{{Laibe}}, \binits{G.}},
\bauthor{\bsnm{{M{\'e}nard}}, \binits{F.}},
\bauthor{\bsnm{{Poblete}}, \binits{P.P.}},
\bauthor{\bsnm{{Montesinos}}, \binits{M.}}:
\batitle{{Flybys in protoplanetary discs: I. Gas and dust dynamics}}.
\bjtitle{\mnras}
\bvolume{483},
\bfpage{4114}--\blpage{4139}
(\byear{2019}).
doi:\doiurl{10.1093/mnras/sty3325}.
\arxivurl{1812.00961}
\end{barticle}
\endbibitem

\bibitem{DESTINY}
\begin{botherref}
\oauthor{\bsnm{DESTINY}}:
http://www3.mpifr-bonn.mpg.de/encounter-properties/
(2019).
Accessed 7 May 2019.
\end{botherref}
\endbibitem

\bibitem{Rosotti2014}
\begin{barticle}
\bauthor{\bsnm{{Rosotti}}, \binits{G.P.}},
\bauthor{\bsnm{{Dale}}, \binits{J.E.}},
\bauthor{\bsnm{{de Juan Ovelar}}, \binits{M.}},
\bauthor{\bsnm{{Hubber}}, \binits{D.A.}},
\bauthor{\bsnm{{Kruijssen}}, \binits{J.M.D.}},
\bauthor{\bsnm{{Ercolano}}, \binits{B.}},
\bauthor{\bsnm{{Walch}}, \binits{S.}}:
\batitle{{Protoplanetary disc evolution affected by star-disc interactions in
  young stellar clusters}}.
\bjtitle{\mnras}
\bvolume{441},
\bfpage{2094}--\blpage{2110}
(\byear{2014}).
doi:\doiurl{10.1093/mnras/stu679}.
\arxivurl{1404.1931}
\end{barticle}
\endbibitem

\bibitem{Vincke2015}
\begin{barticle}
\bauthor{\bsnm{{Vincke}}, \binits{K.}},
\bauthor{\bsnm{{Breslau}}, \binits{A.}},
\bauthor{\bsnm{{Pfalzner}}, \binits{S.}}:
\batitle{{Strong effect of the cluster environment on the size of
  protoplanetary discs?}}
\bjtitle{\aap}
\bvolume{577},
\bfpage{115}
(\byear{2015}).
doi:\doiurl{10.1051/0004-6361/201425552}.
\arxivurl{1504.06092}
\end{barticle}
\endbibitem

\bibitem{Vincke2016}
\begin{barticle}
\bauthor{\bsnm{{Vincke}}, \binits{K.}},
\bauthor{\bsnm{{Pfalzner}}, \binits{S.}}:
\batitle{{Cluster Dynamics Largely Shapes Protoplanetary Disk Sizes}}.
\bjtitle{\apj}
\bvolume{828},
\bfpage{48}
(\byear{2016}).
doi:\doiurl{10.3847/0004-637X/828/1/48}.
\arxivurl{1606.07431}
\end{barticle}
\endbibitem

\bibitem{Zwart2016}
\begin{barticle}
\bauthor{\bsnm{{Portegies Zwart}}, \binits{S.F.}}:
\batitle{{Stellar disc destruction by dynamical interactions in the Orion
  Trapezium star cluster}}.
\bjtitle{\mnras}
\bvolume{457},
\bfpage{313}--\blpage{319}
(\byear{2016}).
doi:\doiurl{10.1093/mnras/stv2831}.
\arxivurl{1511.08900}
\end{barticle}
\endbibitem

\bibitem{Cai2018}
\begin{barticle}
\bauthor{\bsnm{{Cai}}, \binits{M.X.}},
\bauthor{\bsnm{{Portegies Zwart}}, \binits{S.}},
\bauthor{\bsnm{{van Elteren}}, \binits{A.}}:
\batitle{{The signatures of the parental cluster on field planetary systems}}.
\bjtitle{\mnras}
\bvolume{474},
\bfpage{5114}--\blpage{5121}
(\byear{2018}).
doi:\doiurl{10.1093/mnras/stx3064}.
\arxivurl{1711.01274}
\end{barticle}
\endbibitem

\bibitem{Zwart2019}
\begin{barticle}
\bauthor{\bsnm{{Portegies Zwart}}, \binits{S.}}:
\batitle{{The formation of solar-system analogs in young star clusters}}.
\bjtitle{\aap}
\bvolume{622},
\bfpage{69}
(\byear{2019}).
doi:\doiurl{10.1051/0004-6361/201833974}
\end{barticle}
\endbibitem

\bibitem{Vincke2018}
\begin{barticle}
\bauthor{\bsnm{{Vincke}}, \binits{K.}},
\bauthor{\bsnm{{Pfalzner}}, \binits{S.}}:
\batitle{{How Do Disks and Planetary Systems in High-mass Open Clusters Differ
  from Those around Field Stars?}}
\bjtitle{\apj}
\bvolume{868},
\bfpage{1}
(\byear{2018}).
doi:\doiurl{10.3847/1538-4357/aae7d1}.
\arxivurl{1810.04453}
\end{barticle}
\endbibitem

\bibitem{Concha2019}
\begin{barticle}
\bauthor{\bsnm{{Concha-Ram{\'{\i}}rez}}, \binits{F.}},
\bauthor{\bsnm{{Vaher}}, \binits{E.}},
\bauthor{\bsnm{{Portegies Zwart}}, \binits{S.}}:
\batitle{{The viscous evolution of circumstellar discs in young star
  clusters}}.
\bjtitle{\mnras}
\bvolume{482},
\bfpage{732}--\blpage{742}
(\byear{2019}).
doi:\doiurl{10.1093/mnras/sty2721}.
\arxivurl{1810.02368}
\end{barticle}
\endbibitem

\bibitem{Thies2005}
\begin{barticle}
\bauthor{\bsnm{{Thies}}, \binits{I.}},
\bauthor{\bsnm{{Kroupa}}, \binits{P.}},
\bauthor{\bsnm{{Theis}}, \binits{C.}}:
\batitle{{Induced planet formation in stellar clusters: a parameter study of
  star-disc encounters}}.
\bjtitle{\mnras}
\bvolume{364},
\bfpage{961}--\blpage{970}
(\byear{2005}).
doi:\doiurl{10.1111/j.1365-2966.2005.09644.x}.
\arxivurl{astro-ph/0510007}
\end{barticle}
\endbibitem

\bibitem{Jilkova2015}
\begin{barticle}
\bauthor{\bsnm{{J{\'i}lkov{\'a}}}, \binits{L.}},
\bauthor{\bsnm{{Portegies Zwart}}, \binits{S.}},
\bauthor{\bsnm{{Pijloo}}, \binits{T.}},
\bauthor{\bsnm{{Hammer}}, \binits{M.}}:
\batitle{{How Sedna and family were captured in a close encounter with a solar
  sibling}}.
\bjtitle{\mnras}
\bvolume{453},
\bfpage{3157}--\blpage{3162}
(\byear{2015}).
doi:\doiurl{10.1093/mnras/stv1803}.
\arxivurl{1506.03105}
\end{barticle}
\endbibitem

\bibitem{Mustill2016}
\begin{barticle}
\bauthor{\bsnm{{Mustill}}, \binits{A.J.}},
\bauthor{\bsnm{{Raymond}}, \binits{S.N.}},
\bauthor{\bsnm{{Davies}}, \binits{M.B.}}:
\batitle{{Is there an exoplanet in the Solar system?}}
\bjtitle{\mnras}
\bvolume{460},
\bfpage{109}--\blpage{113}
(\byear{2016}).
doi:\doiurl{10.1093/mnrasl/slw075}.
\arxivurl{1603.07247}
\end{barticle}
\endbibitem

\bibitem{Pfalzner2018a}
\begin{barticle}
\bauthor{\bsnm{{Pfalzner}}, \binits{S.}},
\bauthor{\bsnm{{Bhandare}}, \binits{A.}},
\bauthor{\bsnm{{Vincke}}, \binits{K.}}:
\batitle{{Did a stellar fly-by shape the planetary system around Pr 0211 in the
  cluster M44?}}
\bjtitle{\aap}
\bvolume{610},
\bfpage{33}
(\byear{2018}).
doi:\doiurl{10.1051/0004-6361/201731375}.
\arxivurl{1711.06043}
\end{barticle}
\endbibitem

\bibitem{Olczak2006}
\begin{barticle}
\bauthor{\bsnm{{Olczak}}, \binits{C.}},
\bauthor{\bsnm{{Pfalzner}}, \binits{S.}},
\bauthor{\bsnm{{Spurzem}}, \binits{R.}}:
\batitle{{Encounter-triggered Disk Mass Loss in the Orion Nebula Cluster}}.
\bjtitle{\apj}
\bvolume{642},
\bfpage{1140}--\blpage{1151}
(\byear{2006}).
doi:\doiurl{10.1086/501044}.
\arxivurl{astro-ph/0601166}
\end{barticle}
\endbibitem

\bibitem{ovelar2012}
\begin{barticle}
\bauthor{\bsnm{{de Juan Ovelar}}, \binits{M.}},
\bauthor{\bsnm{{Kruijssen}}, \binits{J.M.D.}},
\bauthor{\bsnm{{Bressert}}, \binits{E.}},
\bauthor{\bsnm{{Testi}}, \binits{L.}},
\bauthor{\bsnm{{Bastian}}, \binits{N.}},
\bauthor{\bsnm{{C{\'a}novas}}, \binits{H.}}:
\batitle{{Can habitable planets form in clustered environments?}}
\bjtitle{\aap}
\bvolume{546},
\bfpage{1}
(\byear{2012}).
doi:\doiurl{10.1051/0004-6361/201219627}.
\arxivurl{1209.2136}
\end{barticle}
\endbibitem

\bibitem{Pringle1981}
\begin{barticle}
\bauthor{\bsnm{{Pringle}}, \binits{J.E.}}:
\batitle{{Accretion discs in astrophysics}}.
\bjtitle{\araa}
\bvolume{19},
\bfpage{137}--\blpage{162}
(\byear{1981}).
doi:\doiurl{10.1146/annurev.aa.19.090181.001033}
\end{barticle}
\endbibitem

\bibitem{Pfalzner2003}
\begin{barticle}
\bauthor{\bsnm{{Pfalzner}}, \binits{S.}}:
\batitle{{Spiral Arms in Accretion Disk Encounters}}.
\bjtitle{\apj}
\bvolume{592},
\bfpage{986}--\blpage{1001}
(\byear{2003}).
doi:\doiurl{10.1086/375808}
\end{barticle}
\endbibitem

\bibitem{Andrews2013}
\begin{barticle}
\bauthor{\bsnm{{Andrews}}, \binits{S.M.}},
\bauthor{\bsnm{{Rosenfeld}}, \binits{K.A.}},
\bauthor{\bsnm{{Kraus}}, \binits{A.L.}},
\bauthor{\bsnm{{Wilner}}, \binits{D.J.}}:
\batitle{{The Mass Dependence between Protoplanetary Disks and their Stellar
  Hosts}}.
\bjtitle{\apj}
\bvolume{771},
\bfpage{129}
(\byear{2013}).
doi:\doiurl{10.1088/0004-637X/771/2/129}.
\arxivurl{1305.5262}
\end{barticle}
\endbibitem

\bibitem{Musielak2014}
\begin{barticle}
\bauthor{\bsnm{{Musielak}}, \binits{Z.E.}},
\bauthor{\bsnm{{Quarles}}, \binits{B.}}:
\batitle{{The three-body problem}}.
\bjtitle{Reports on Progress in Physics}
\bvolume{77}(\bissue{6}),
\bfpage{065901}
(\byear{2014}).
doi:\doiurl{10.1088/0034-4885/77/6/065901}.
\arxivurl{1508.02312}
\end{barticle}
\endbibitem

\bibitem{PfalznerUmbreit2005}
\begin{barticle}
\bauthor{\bsnm{{Pfalzner}}, \binits{S.}},
\bauthor{\bsnm{{Umbreit}}, \binits{S.}},
\bauthor{\bsnm{{Henning}}, \binits{T.}}:
\batitle{{Disk-Disk Encounters between Low-Mass Protoplanetary Accretion
  Disks}}.
\bjtitle{\apj}
\bvolume{629},
\bfpage{526}--\blpage{534}
(\byear{2005}).
doi:\doiurl{10.1086/431350}.
\arxivurl{astro-ph/0504590}
\end{barticle}
\endbibitem

\bibitem{Weidner2010}
\begin{barticle}
\bauthor{\bsnm{{Weidner}}, \binits{C.}},
\bauthor{\bsnm{{Kroupa}}, \binits{P.}},
\bauthor{\bsnm{{Bonnell}}, \binits{I.A.D.}}:
\batitle{{The relation between the most-massive star and its parental star
  cluster mass}}.
\bjtitle{\mnras}
\bvolume{401},
\bfpage{275}--\blpage{293}
(\byear{2010}).
doi:\doiurl{10.1111/j.1365-2966.2009.15633.x}.
\arxivurl{0909.1555}
\end{barticle}
\endbibitem

\bibitem{Olczak2012}
\begin{barticle}
\bauthor{\bsnm{{Olczak}}, \binits{C.}},
\bauthor{\bsnm{{Kaczmarek}}, \binits{T.}},
\bauthor{\bsnm{{Harfst}}, \binits{S.}},
\bauthor{\bsnm{{Pfalzner}}, \binits{S.}},
\bauthor{\bsnm{{Portegies Zwart}}, \binits{S.}}:
\batitle{{The Evolution of Protoplanetary Disks in the Arches Cluster}}.
\bjtitle{\apj}
\bvolume{756}(\bissue{2}),
\bfpage{123}
(\byear{2012}).
doi:\doiurl{10.1088/0004-637X/756/2/123}.
\arxivurl{1207.2256}
\end{barticle}
\endbibitem

\bibitem{Steinhausen2012}
\begin{barticle}
\bauthor{\bsnm{{Steinhausen}}, \binits{M.}},
\bauthor{\bsnm{{Olczak}}, \binits{C.}},
\bauthor{\bsnm{{Pfalzner}}, \binits{S.}}:
\batitle{{Disc-mass distribution in star-disc encounters}}.
\bjtitle{\aap}
\bvolume{538},
\bfpage{10}
(\byear{2012}).
doi:\doiurl{10.1051/0004-6361/201117682}.
\arxivurl{1111.2466}
\end{barticle}
\endbibitem

\bibitem{Boffin1998}
\begin{barticle}
\bauthor{\bsnm{{Boffin}}, \binits{H.M.J.}},
\bauthor{\bsnm{{Watkins}}, \binits{S.J.}},
\bauthor{\bsnm{{Bhattal}}, \binits{A.S.}},
\bauthor{\bsnm{{Francis}}, \binits{N.}},
\bauthor{\bsnm{{Whitworth}}, \binits{A.P.}}:
\batitle{{Numerical simulations of protostellar encounters - I. Star-disc
  encounters}}.
\bjtitle{\mnras}
\bvolume{300},
\bfpage{1189}--\blpage{1204}
(\byear{1998}).
doi:\doiurl{10.1046/j.1365-8711.1998.01986.x}.
\arxivurl{astro-ph/9805349}
\end{barticle}
\endbibitem

\bibitem{Pfalzner2004}
\begin{barticle}
\bauthor{\bsnm{{Pfalzner}}, \binits{S.}}:
\batitle{{Angular Momentum Transfer in Star-Disk Encounters: The Case of
  Low-Mass Disks}}.
\bjtitle{\apj}
\bvolume{602},
\bfpage{356}--\blpage{362}
(\byear{2004}).
doi:\doiurl{10.1086/381023}.
\arxivurl{astro-ph/0310743}
\end{barticle}
\endbibitem

\bibitem{Pfalzner2007}
\begin{barticle}
\bauthor{\bsnm{{Pfalzner}}, \binits{S.}},
\bauthor{\bsnm{{Olczak}}, \binits{C.}}:
\batitle{{Gravitational instabilities induced by cluster environment? The
  encounter-induced angular momentum transfer in discs}}.
\bjtitle{\aap}
\bvolume{462},
\bfpage{193}--\blpage{198}
(\byear{2007}).
doi:\doiurl{10.1051/0004-6361:20066037}.
\arxivurl{astro-ph/0609519}
\end{barticle}
\endbibitem

\bibitem{Dai2015}
\begin{barticle}
\bauthor{\bsnm{{Dai}}, \binits{F.}},
\bauthor{\bsnm{{Facchini}}, \binits{S.}},
\bauthor{\bsnm{{Clarke}}, \binits{C.J.}},
\bauthor{\bsnm{{Haworth}}, \binits{T.J.}}:
\batitle{{A tidal encounter caught in the act: modelling a star-disc fly-by in
  the young RW Aurigae system}}.
\bjtitle{\mnras}
\bvolume{449},
\bfpage{1996}--\blpage{2009}
(\byear{2015}).
doi:\doiurl{10.1093/mnras/stv403}.
\arxivurl{1502.06649}
\end{barticle}
\endbibitem

\bibitem{Xiang2016}
\begin{barticle}
\bauthor{\bsnm{{Xiang-Gruess}}, \binits{M.}}:
\batitle{{Generation of highly inclined protoplanetary discs through single
  stellar flybys}}.
\bjtitle{\mnras}
\bvolume{455},
\bfpage{3086}--\blpage{3100}
(\byear{2016}).
doi:\doiurl{10.1093/mnras/stv2514}.
\arxivurl{1510.07458}
\end{barticle}
\endbibitem

\bibitem{Pfalzner2018}
\begin{barticle}
\bauthor{\bsnm{{Pfalzner}}, \binits{S.}},
\bauthor{\bsnm{{Bhandare}}, \binits{A.}},
\bauthor{\bsnm{{Vincke}}, \binits{K.}},
\bauthor{\bsnm{{Lacerda}}, \binits{P.}}:
\batitle{{Outer Solar System Possibly Shaped by a Stellar Fly-by}}.
\bjtitle{\apj}
\bvolume{863},
\bfpage{45}
(\byear{2018}).
doi:\doiurl{10.3847/1538-4357/aad23c}.
\arxivurl{1807.02960}
\end{barticle}
\endbibitem

\bibitem{Zheng2015}
\begin{barticle}
\bauthor{\bsnm{{Zheng}}, \binits{X.}},
\bauthor{\bsnm{{Kouwenhoven}}, \binits{M.B.N.}},
\bauthor{\bsnm{{Wang}}, \binits{L.}}:
\batitle{{The dynamical fate of planetary systems in young star clusters}}.
\bjtitle{\mnras}
\bvolume{453}(\bissue{3}),
\bfpage{2759}--\blpage{2770}
(\byear{2015}).
doi:\doiurl{10.1093/mnras/stv1832}.
\arxivurl{1508.01593}
\end{barticle}
\endbibitem

\bibitem{Dotti2018}
\begin{botherref}
\oauthor{\bsnm{{Flammini Dotti}}, \binits{F.}},
\oauthor{\bsnm{{Cai}}, \binits{M.X.}},
\oauthor{\bsnm{{Spurzem}}, \binits{R.}},
\oauthor{\bsnm{{Kouwenhoven}}, \binits{M.B.N.}}:
{Planetary Systems in Star Clusters: the dynamical evolution and survival}.
arXiv e-prints,
1811--12660
(2018).
\arxivurl{1811.12660}
\end{botherref}
\endbibitem

\bibitem{Fujii2019}
\begin{barticle}
\bauthor{\bsnm{{Fujii}}, \binits{M.S.}},
\bauthor{\bsnm{{Hori}}, \binits{Y.}}:
\batitle{{Survival rates of planets in open clusters: the Pleiades, Hyades, and
  Praesepe clusters}}.
\bjtitle{\aap}
\bvolume{624},
\bfpage{110}
(\byear{2019}).
doi:\doiurl{10.1051/0004-6361/201834677}.
\arxivurl{1811.08598}
\end{barticle}
\endbibitem

\end{thebibliography}

\newcommand{\BMCxmlcomment}[1]{}

\BMCxmlcomment{

<refgrp>

<bibl id="B1">
  <title><p>{ALMA Continuum Observations of a 30 Myr Old Gaseous Debris Disk
  around HD 21997}</p></title>
  <aug>
    <au><snm>{Mo{\'o}r}</snm><fnm>A.</fnm></au>
    <au><snm>{Juh{\'a}sz}</snm><fnm>A.</fnm></au>
    <au><snm>{K{\'o}sp{\'a}l}</snm><fnm>{\'A}.</fnm></au>
    <au><snm>{{\'A}brah{\'a}m}</snm><fnm>P.</fnm></au>
    <au><snm>{Apai}</snm><fnm>D.</fnm></au>
    <au><snm>{Csengeri}</snm><fnm>T.</fnm></au>
    <au><snm>{Grady}</snm><fnm>C.</fnm></au>
    <au><snm>{Henning}</snm><fnm>T.</fnm></au>
    <au><snm>{Hughes}</snm><fnm>A. M.</fnm></au>
    <au><snm>{Kiss}</snm><fnm>C.</fnm></au>
    <au><snm>{Pascucci}</snm><fnm>I.</fnm></au>
    <au><snm>{Schmalzl}</snm><fnm>M.</fnm></au>
    <au><snm>{Gab{\'a}nyi}</snm><fnm>K.</fnm></au>
  </aug>
  <source>\apjl</source>
  <pubdate>2013</pubdate>
  <volume>777</volume>
  <fpage>L25</fpage>
</bibl>

<bibl id="B2">
  <title><p>{ALMA Observations of the Orion Proplyds}</p></title>
  <aug>
    <au><snm>{Mann}</snm><fnm>R. K.</fnm></au>
    <au><snm>{Di Francesco}</snm><fnm>J.</fnm></au>
    <au><snm>{Johnstone}</snm><fnm>D.</fnm></au>
    <au><snm>{Andrews}</snm><fnm>S. M.</fnm></au>
    <au><snm>{Williams}</snm><fnm>J. P.</fnm></au>
    <au><snm>{Bally}</snm><fnm>J.</fnm></au>
    <au><snm>{Ricci}</snm><fnm>L.</fnm></au>
    <au><snm>{Hughes}</snm><fnm>A. M.</fnm></au>
    <au><snm>{Matthews}</snm><fnm>B. C.</fnm></au>
  </aug>
  <source>\apj</source>
  <pubdate>2014</pubdate>
  <volume>784</volume>
  <fpage>82</fpage>
</bibl>

<bibl id="B3">
  <title><p>{ALMA Observations of the Largest Proto-Planetary Disk in the Orion
  Nebula, 114-426: A CO Silhouette}</p></title>
  <aug>
    <au><snm>{Bally}</snm><fnm>J.</fnm></au>
    <au><snm>{Mann}</snm><fnm>R. K.</fnm></au>
    <au><snm>{Eisner}</snm><fnm>J.</fnm></au>
    <au><snm>{Andrews}</snm><fnm>S. M.</fnm></au>
    <au><snm>{Di Francesco}</snm><fnm>J.</fnm></au>
    <au><snm>{Hughes}</snm><fnm>M.</fnm></au>
    <au><snm>{Johnstone}</snm><fnm>D.</fnm></au>
    <au><snm>{Matthews}</snm><fnm>B.</fnm></au>
    <au><snm>{Ricci}</snm><fnm>L.</fnm></au>
    <au><snm>{Williams}</snm><fnm>J. P.</fnm></au>
  </aug>
  <source>\apj</source>
  <pubdate>2015</pubdate>
  <volume>808</volume>
  <fpage>69</fpage>
</bibl>

<bibl id="B4">
  <title><p>{Revolutionizing our View of Protostellar Multiplicity and Disks:
  The VLA Nascent Disk and Multiplicity (VANDAM) Survey of the Perseus
  Molecular Cloud}</p></title>
  <aug>
    <au><snm>{Tobin}</snm><fnm>J. J.</fnm></au>
    <au><snm>{Looney}</snm><fnm>L. W.</fnm></au>
    <au><snm>{Li}</snm><fnm>Z. Y.</fnm></au>
    <au><snm>{Chandler}</snm><fnm>C. J.</fnm></au>
    <au><snm>{Dunham}</snm><fnm>M. M.</fnm></au>
    <au><snm>{Segura-Cox}</snm><fnm>D.</fnm></au>
    <au><snm>{Cox}</snm><fnm>E. G.</fnm></au>
    <au><snm>{Harris}</snm><fnm>R. J.</fnm></au>
    <au><snm>{Melis}</snm><fnm>C.</fnm></au>
    <au><snm>{Sadavoy}</snm><fnm>S. I.</fnm></au>
    <au><snm>{P{\'e}rez}</snm><fnm>L.</fnm></au>
    <au><snm>{Kratter}</snm><fnm>K.</fnm></au>
  </aug>
  <source>EAS Publications Series</source>
  <series><title><p>EAS Publications Series</p></title></series>
  <pubdate>2015</pubdate>
  <volume>75-76</volume>
  <fpage>273</fpage>
  <lpage>276</lpage>
</bibl>

<bibl id="B5">
  <title><p>{The Disk Substructures at High Angular Resolution Project
  (DSHARP). I. Motivation, Sample, Calibration, and Overview}</p></title>
  <aug>
    <au><snm>{Andrews}</snm><fnm>SM</fnm></au>
    <au><snm>{Huang}</snm><fnm>J</fnm></au>
    <au><snm>{P{\'e}rez}</snm><fnm>LM</fnm></au>
    <au><snm>{Isella}</snm><fnm>A</fnm></au>
    <au><snm>{Dullemond}</snm><fnm>CP</fnm></au>
    <au><snm>{Kurtovic}</snm><fnm>NT</fnm></au>
    <au><snm>{Guzm{\'a}n}</snm><fnm>VV</fnm></au>
    <au><snm>{Carpenter}</snm><fnm>JM</fnm></au>
    <au><snm>{Wilner}</snm><fnm>DJ</fnm></au>
    <au><snm>{Zhang}</snm><fnm>S</fnm></au>
    <au><snm>{Zhu}</snm><fnm>Z</fnm></au>
    <au><snm>{Birnstiel}</snm><fnm>T</fnm></au>
    <au><snm>{Bai}</snm><fnm>XN</fnm></au>
    <au><snm>{Benisty}</snm><fnm>M</fnm></au>
    <au><snm>{Hughes}</snm><fnm>AM</fnm></au>
    <au><snm>{{\"O}berg}</snm><fnm>KI</fnm></au>
    <au><snm>{Ricci}</snm><fnm>L</fnm></au>
  </aug>
  <source>\apj</source>
  <pubdate>2018</pubdate>
  <volume>869</volume>
  <fpage>L41</fpage>
</bibl>

<bibl id="B6">
  <title><p>{A New Hybrid N-body-coagulation Code for the Formation of Gas
  Giant Planets}</p></title>
  <aug>
    <au><snm>{Bromley}</snm><fnm>BC</fnm></au>
    <au><snm>{Kenyon}</snm><fnm>SJ</fnm></au>
  </aug>
  <source>\apj</source>
  <pubdate>2011</pubdate>
  <volume>731</volume>
  <fpage>101</fpage>
</bibl>

<bibl id="B7">
  <title><p>{Planet-Disk Interactions and Early Evolution of Planetary
  Systems}</p></title>
  <aug>
    <au><snm>{Baruteau}</snm><fnm>C.</fnm></au>
    <au><snm>{Crida}</snm><fnm>A.</fnm></au>
    <au><snm>{Paardekooper}</snm><fnm>S. J.</fnm></au>
    <au><snm>{Masset}</snm><fnm>F.</fnm></au>
    <au><snm>{Guilet}</snm><fnm>J.</fnm></au>
    <au><snm>{Bitsch}</snm><fnm>B.</fnm></au>
    <au><snm>{Nelson}</snm><fnm>R.</fnm></au>
    <au><snm>{Kley}</snm><fnm>W.</fnm></au>
    <au><snm>{Papaloizou}</snm><fnm>J.</fnm></au>
  </aug>
  <source>Protostars and Planets VI</source>
  <editor>{Beuther}, Henrik and {Klessen}, Ralf S. and {Dullemond}, Cornelis P.
  and {Henning}, Thomas</editor>
  <pubdate>2014</pubdate>
  <fpage>667</fpage>
</bibl>

<bibl id="B8">
  <title><p>{The structure of protoplanetary discs around evolving young
  stars}</p></title>
  <aug>
    <au><snm>{Bitsch}</snm><fnm>B</fnm></au>
    <au><snm>{Johansen}</snm><fnm>A</fnm></au>
    <au><snm>{Lambrechts}</snm><fnm>M</fnm></au>
    <au><snm>{Morbidelli}</snm><fnm>A</fnm></au>
  </aug>
  <source>\aap</source>
  <pubdate>2015</pubdate>
  <volume>575</volume>
  <fpage>A28</fpage>
</bibl>

<bibl id="B9">
  <title><p>{Planet Formation around Binary Stars: Tatooine Made
  Easy}</p></title>
  <aug>
    <au><snm>{Bromley}</snm><fnm>BC</fnm></au>
    <au><snm>{Kenyon}</snm><fnm>SJ</fnm></au>
  </aug>
  <source>\apj</source>
  <pubdate>2015</pubdate>
  <volume>806</volume>
  <fpage>98</fpage>
</bibl>

<bibl id="B10">
  <title><p>{The Formation of Stellar Clusters}</p></title>
  <aug>
    <au><snm>{Clarke}</snm><fnm>C. J.</fnm></au>
    <au><snm>{Bonnell}</snm><fnm>I. A.</fnm></au>
    <au><snm>{Hillenbrand}</snm><fnm>L. A.</fnm></au>
  </aug>
  <source>Protostars and Planets IV</source>
  <pubdate>2000</pubdate>
  <fpage>151</fpage>
</bibl>

<bibl id="B11">
  <title><p>{Embedded Clusters in Molecular Clouds}</p></title>
  <aug>
    <au><snm>{Lada}</snm><fnm>C. J.</fnm></au>
    <au><snm>{Lada}</snm><fnm>E. A.</fnm></au>
  </aug>
  <source>\araa</source>
  <pubdate>2003</pubdate>
  <volume>41</volume>
  <fpage>57</fpage>
  <lpage>-115</lpage>
</bibl>

<bibl id="B12">
  <title><p>{A Catalog of Young Stellar Groups and Clusters within 1 Kiloparsec
  of the Sun}</p></title>
  <aug>
    <au><snm>{Porras}</snm><fnm>A.</fnm></au>
    <au><snm>{Christopher}</snm><fnm>M.</fnm></au>
    <au><snm>{Allen}</snm><fnm>L.</fnm></au>
    <au><snm>{Di Francesco}</snm><fnm>J.</fnm></au>
    <au><snm>{Megeath}</snm><fnm>S. T.</fnm></au>
    <au><snm>{Myers}</snm><fnm>P. C.</fnm></au>
  </aug>
  <source>\aj</source>
  <pubdate>2003</pubdate>
  <volume>126</volume>
  <fpage>1916</fpage>
  <lpage>-1924</lpage>
</bibl>

<bibl id="B13">
  <title><p>{Disk Dispersal around Young Stars}</p></title>
  <aug>
    <au><snm>{Hollenbach}</snm><fnm>D. J.</fnm></au>
    <au><snm>{Yorke}</snm><fnm>H. W.</fnm></au>
    <au><snm>{Johnstone}</snm><fnm>D.</fnm></au>
  </aug>
  <source>Protostars and Planets IV</source>
  <pubdate>2000</pubdate>
  <fpage>401</fpage>
</bibl>

<bibl id="B14">
  <title><p>{Protoplanetary Disks and Their Evolution}</p></title>
  <aug>
    <au><snm>{Williams}</snm><fnm>J. P.</fnm></au>
    <au><snm>{Cieza}</snm><fnm>L. A.</fnm></au>
  </aug>
  <source>\araa</source>
  <pubdate>2011</pubdate>
  <volume>49</volume>
  <fpage>67</fpage>
  <lpage>-117</lpage>
</bibl>

<bibl id="B15">
  <title><p>{Photoevaporation of Disks and Clumps by Nearby Massive Stars:
  Application to Disk Destruction in the Orion Nebula}</p></title>
  <aug>
    <au><snm>{Johnstone}</snm><fnm>D.</fnm></au>
    <au><snm>{Hollenbach}</snm><fnm>D.</fnm></au>
    <au><snm>{Bally}</snm><fnm>J.</fnm></au>
  </aug>
  <source>\apj</source>
  <pubdate>1998</pubdate>
  <volume>499</volume>
  <fpage>758</fpage>
  <lpage>-776</lpage>
</bibl>

<bibl id="B16">
  <title><p>{Photoevaporation of Circumstellar Disks Due to External
  Far-Ultraviolet Radiation in Stellar Aggregates}</p></title>
  <aug>
    <au><snm>{Adams}</snm><fnm>F. C.</fnm></au>
    <au><snm>{Hollenbach}</snm><fnm>D.</fnm></au>
    <au><snm>{Laughlin}</snm><fnm>G.</fnm></au>
    <au><snm>{Gorti}</snm><fnm>U.</fnm></au>
  </aug>
  <source>\apj</source>
  <pubdate>2004</pubdate>
  <volume>611</volume>
  <fpage>360</fpage>
  <lpage>-379</lpage>
</bibl>

<bibl id="B17">
  <title><p>{Photoevaporation of Circumstellar Disks around Young
  Stars}</p></title>
  <aug>
    <au><snm>{Font}</snm><fnm>A. S.</fnm></au>
    <au><snm>{McCarthy}</snm><fnm>I. G.</fnm></au>
    <au><snm>{Johnstone}</snm><fnm>D.</fnm></au>
    <au><snm>{Ballantyne}</snm><fnm>D. R.</fnm></au>
  </aug>
  <source>\apj</source>
  <pubdate>2004</pubdate>
  <volume>607</volume>
  <fpage>890</fpage>
  <lpage>-903</lpage>
</bibl>

<bibl id="B18">
  <title><p>{The photoevaporation of discs around young stars in massive
  clusters}</p></title>
  <aug>
    <au><snm>{Clarke}</snm><fnm>C. J.</fnm></au>
  </aug>
  <source>\mnras</source>
  <pubdate>2007</pubdate>
  <volume>376</volume>
  <fpage>1350</fpage>
  <lpage>-1356</lpage>
</bibl>

<bibl id="B19">
  <title><p>{Models of the Structure and Evolution of Protoplanetary
  Disks}</p></title>
  <aug>
    <au><snm>{Dullemond}</snm><fnm>C. P.</fnm></au>
    <au><snm>{Hollenbach}</snm><fnm>D.</fnm></au>
    <au><snm>{Kamp}</snm><fnm>I.</fnm></au>
    <au><snm>{D'Alessio}</snm><fnm>P.</fnm></au>
  </aug>
  <source>Protostars and Planets V</source>
  <pubdate>2007</pubdate>
  <fpage>555</fpage>
  <lpage>-572</lpage>
</bibl>

<bibl id="B20">
  <title><p>{Photoevaporation of Circumstellar Disks By Far-Ultraviolet,
  Extreme-Ultraviolet and X-Ray Radiation from the Central Star}</p></title>
  <aug>
    <au><snm>{Gorti}</snm><fnm>U.</fnm></au>
    <au><snm>{Hollenbach}</snm><fnm>D.</fnm></au>
  </aug>
  <source>\apj</source>
  <pubdate>2009</pubdate>
  <volume>690</volume>
  <fpage>1539</fpage>
  <lpage>-1552</lpage>
</bibl>

<bibl id="B21">
  <title><p>{Radiation-hydrodynamic models of X-ray and EUV photoevaporating
  protoplanetary discs}</p></title>
  <aug>
    <au><snm>{Owen}</snm><fnm>J. E.</fnm></au>
    <au><snm>{Ercolano}</snm><fnm>B.</fnm></au>
    <au><snm>{Clarke}</snm><fnm>C. J.</fnm></au>
    <au><snm>{Alexander}</snm><fnm>R. D.</fnm></au>
  </aug>
  <source>\mnras</source>
  <pubdate>2010</pubdate>
  <volume>401</volume>
  <fpage>1415</fpage>
  <lpage>-1428</lpage>
</bibl>

<bibl id="B22">
  <title><p>{On the theory of disc photoevaporation}</p></title>
  <aug>
    <au><snm>{Owen}</snm><fnm>J. E.</fnm></au>
    <au><snm>{Clarke}</snm><fnm>C. J.</fnm></au>
    <au><snm>{Ercolano}</snm><fnm>B.</fnm></au>
  </aug>
  <source>\mnras</source>
  <pubdate>2012</pubdate>
  <volume>422</volume>
  <fpage>1880</fpage>
  <lpage>-1901</lpage>
</bibl>

<bibl id="B23">
  <title><p>{The long-term evolution of photoevaporating transition discs with
  giant planets}</p></title>
  <aug>
    <au><snm>{Rosotti}</snm><fnm>G. P.</fnm></au>
    <au><snm>{Ercolano}</snm><fnm>B.</fnm></au>
    <au><snm>{Owen}</snm><fnm>J. E.</fnm></au>
  </aug>
  <source>\mnras</source>
  <pubdate>2015</pubdate>
  <volume>454</volume>
  <fpage>2173</fpage>
  <lpage>-2182</lpage>
</bibl>

<bibl id="B24">
  <title><p>{Disc truncation in embedded star clusters: Dynamical encounters
  versus face-on accretion}</p></title>
  <aug>
    <au><snm>{Wijnen}</snm><fnm>T. P. G.</fnm></au>
    <au><snm>{Pols}</snm><fnm>O. R.</fnm></au>
    <au><snm>{Pelupessy}</snm><fnm>F. I.</fnm></au>
    <au><snm>{Portegies Zwart}</snm><fnm>S.</fnm></au>
  </aug>
  <source>\aap</source>
  <pubdate>2017</pubdate>
  <volume>604</volume>
  <fpage>A91</fpage>
</bibl>

<bibl id="B25">
  <title><p>{Encounters with Protostellar Disks. II. Disruption and Binary
  Formation}</p></title>
  <aug>
    <au><snm>{Heller}</snm><fnm>C. H.</fnm></au>
  </aug>
  <source>\apj</source>
  <pubdate>1995</pubdate>
  <volume>455</volume>
  <fpage>252</fpage>
</bibl>

<bibl id="B26">
  <title><p>{Energetics of star-disc encounters in the non-linear
  regime}</p></title>
  <aug>
    <au><snm>{Hall}</snm><fnm>S. M.</fnm></au>
    <au><snm>{Clarke}</snm><fnm>C. J.</fnm></au>
    <au><snm>{Pringle}</snm><fnm>J. E.</fnm></au>
  </aug>
  <source>\mnras</source>
  <pubdate>1996</pubdate>
  <volume>278</volume>
  <fpage>303</fpage>
  <lpage>-320</lpage>
</bibl>

<bibl id="B27">
  <title><p>{Accretion disc response to a stellar fly-by}</p></title>
  <aug>
    <au><snm>{Clarke}</snm><fnm>C. J.</fnm></au>
    <au><snm>{Pringle}</snm><fnm>J. E.</fnm></au>
  </aug>
  <source>\mnras</source>
  <pubdate>1993</pubdate>
  <volume>261</volume>
  <fpage>190</fpage>
  <lpage>-202</lpage>
</bibl>

<bibl id="B28">
  <title><p>{Parameter study of star-disc encounters}</p></title>
  <aug>
    <au><snm>{Pfalzner}</snm><fnm>S.</fnm></au>
    <au><snm>{Vogel}</snm><fnm>P.</fnm></au>
    <au><snm>{Scharw{\"a}chter}</snm><fnm>J.</fnm></au>
    <au><snm>{Olczak}</snm><fnm>C.</fnm></au>
  </aug>
  <source>\aap</source>
  <pubdate>2005</pubdate>
  <volume>437</volume>
  <fpage>967</fpage>
  <lpage>-976</lpage>
</bibl>

<bibl id="B29">
  <title><p>{The Effects of a Stellar Encounter on a Planetesimal
  Disk}</p></title>
  <aug>
    <au><snm>{Kobayashi}</snm><fnm>H.</fnm></au>
    <au><snm>{Ida}</snm><fnm>S.</fnm></au>
  </aug>
  <source>\icarus</source>
  <pubdate>2001</pubdate>
  <volume>153</volume>
  <fpage>416</fpage>
  <lpage>-429</lpage>
</bibl>

<bibl id="B30">
  <title><p>{The evidence of an early stellar encounter in Edgeworth Kuiper
  belt}</p></title>
  <aug>
    <au><snm>{Kobayashi}</snm><fnm>H.</fnm></au>
    <au><snm>{Ida}</snm><fnm>S.</fnm></au>
    <au><snm>{Tanaka}</snm><fnm>H.</fnm></au>
  </aug>
  <source>\icarus</source>
  <pubdate>2005</pubdate>
  <volume>177</volume>
  <fpage>246</fpage>
  <lpage>-255</lpage>
</bibl>

<bibl id="B31">
  <title><p>{Sizes of protoplanetary discs after star-disc
  encounters}</p></title>
  <aug>
    <au><snm>{Breslau}</snm><fnm>A.</fnm></au>
    <au><snm>{Steinhausen}</snm><fnm>M.</fnm></au>
    <au><snm>{Vincke}</snm><fnm>K.</fnm></au>
    <au><snm>{Pfalzner}</snm><fnm>S.</fnm></au>
  </aug>
  <source>\aap</source>
  <pubdate>2014</pubdate>
  <volume>565</volume>
  <fpage>A130</fpage>
</bibl>

<bibl id="B32">
  <title><p>{Mass transfer between debris discs during close stellar
  encounters}</p></title>
  <aug>
    <au><snm>{J{\'{\i}}lkov{\'a}}</snm><fnm>L.</fnm></au>
    <au><snm>{Hamers}</snm><fnm>A. S.</fnm></au>
    <au><snm>{Hammer}</snm><fnm>M.</fnm></au>
    <au><snm>{Portegies Zwart}</snm><fnm>S.</fnm></au>
  </aug>
  <source>\mnras</source>
  <pubdate>2016</pubdate>
  <volume>457</volume>
  <fpage>4218</fpage>
  <lpage>4235</lpage>
</bibl>

<bibl id="B33">
  <title><p>{Effects of inclined star-disk encounter on protoplanetary disk
  size}</p></title>
  <aug>
    <au><snm>{Bhandare}</snm><fnm>A.</fnm></au>
    <au><snm>{Breslau}</snm><fnm>A.</fnm></au>
    <au><snm>{Pfalzner}</snm><fnm>S.</fnm></au>
  </aug>
  <source>\aap</source>
  <pubdate>2016</pubdate>
  <volume>594</volume>
  <fpage>A53</fpage>
</bibl>

<bibl id="B34">
  <title><p>{Protoplanetary disc response to distant tidal encounters in
  stellar clusters}</p></title>
  <aug>
    <au><snm>{Winter}</snm><fnm>A. J.</fnm></au>
    <au><snm>{Clarke}</snm><fnm>C. J.</fnm></au>
    <au><snm>{Rosotti}</snm><fnm>G.</fnm></au>
    <au><snm>{Booth}</snm><fnm>R. A.</fnm></au>
  </aug>
  <source>\mnras</source>
  <pubdate>2018</pubdate>
  <volume>475</volume>
  <fpage>2314</fpage>
  <lpage>2325</lpage>
</bibl>

<bibl id="B35">
  <title><p>{Protoplanetary disc truncation mechanisms in stellar clusters:
  comparing external photoevaporation and tidal encounters}</p></title>
  <aug>
    <au><snm>{Winter}</snm><fnm>A. J.</fnm></au>
    <au><snm>{Clarke}</snm><fnm>C. J.</fnm></au>
    <au><snm>{Rosotti}</snm><fnm>G.</fnm></au>
    <au><snm>{Ih}</snm><fnm>J.</fnm></au>
    <au><snm>{Facchini}</snm><fnm>S.</fnm></au>
    <au><snm>{Haworth}</snm><fnm>T. J.</fnm></au>
  </aug>
  <source>\mnras</source>
  <pubdate>2018</pubdate>
  <volume>478</volume>
  <fpage>2700</fpage>
  <lpage>2722</lpage>
</bibl>

<bibl id="B36">
  <title><p>{Flybys in protoplanetary discs: I. Gas and dust
  dynamics}</p></title>
  <aug>
    <au><snm>{Cuello}</snm><fnm>N</fnm></au>
    <au><snm>{Dipierro}</snm><fnm>G</fnm></au>
    <au><snm>{Mentiplay}</snm><fnm>D</fnm></au>
    <au><snm>{Price}</snm><fnm>DJ</fnm></au>
    <au><snm>{Pinte}</snm><fnm>C</fnm></au>
    <au><snm>{Cuadra}</snm><fnm>J</fnm></au>
    <au><snm>{Laibe}</snm><fnm>G</fnm></au>
    <au><snm>{M{\'e}nard}</snm><fnm>F</fnm></au>
    <au><snm>{Poblete}</snm><fnm>PP</fnm></au>
    <au><snm>{Montesinos}</snm><fnm>M</fnm></au>
  </aug>
  <source>\mnras</source>
  <pubdate>2019</pubdate>
  <volume>483</volume>
  <fpage>4114</fpage>
  <lpage>4139</lpage>
</bibl>

<bibl id="B37">
  <title><p>http://www3.mpifr-bonn.mpg.de/encounter-properties/</p></title>
  <aug>
    <au><cnm>DESTINY</cnm></au>
  </aug>
  <pubdate>2019</pubdate>
  <url>http://www3.mpifr-bonn.mpg.de/encounter-properties/</url>
  <note>Accessed 7 May 2019.</note>
</bibl>

<bibl id="B38">
  <title><p>{Protoplanetary disc evolution affected by star-disc interactions
  in young stellar clusters}</p></title>
  <aug>
    <au><snm>{Rosotti}</snm><fnm>G. P.</fnm></au>
    <au><snm>{Dale}</snm><fnm>J. E.</fnm></au>
    <au><snm>{de Juan Ovelar}</snm><fnm>M.</fnm></au>
    <au><snm>{Hubber}</snm><fnm>D. A.</fnm></au>
    <au><snm>{Kruijssen}</snm><fnm>J. M. D.</fnm></au>
    <au><snm>{Ercolano}</snm><fnm>B.</fnm></au>
    <au><snm>{Walch}</snm><fnm>S.</fnm></au>
  </aug>
  <source>\mnras</source>
  <pubdate>2014</pubdate>
  <volume>441</volume>
  <fpage>2094</fpage>
  <lpage>-2110</lpage>
</bibl>

<bibl id="B39">
  <title><p>{Strong effect of the cluster environment on the size of
  protoplanetary discs?}</p></title>
  <aug>
    <au><snm>{Vincke}</snm><fnm>K.</fnm></au>
    <au><snm>{Breslau}</snm><fnm>A.</fnm></au>
    <au><snm>{Pfalzner}</snm><fnm>S.</fnm></au>
  </aug>
  <source>\aap</source>
  <pubdate>2015</pubdate>
  <volume>577</volume>
  <fpage>A115</fpage>
</bibl>

<bibl id="B40">
  <title><p>{Cluster Dynamics Largely Shapes Protoplanetary Disk
  Sizes}</p></title>
  <aug>
    <au><snm>{Vincke}</snm><fnm>K.</fnm></au>
    <au><snm>{Pfalzner}</snm><fnm>S.</fnm></au>
  </aug>
  <source>\apj</source>
  <pubdate>2016</pubdate>
  <volume>828</volume>
  <fpage>48</fpage>
</bibl>

<bibl id="B41">
  <title><p>{Stellar disc destruction by dynamical interactions in the Orion
  Trapezium star cluster}</p></title>
  <aug>
    <au><snm>{Portegies Zwart}</snm><fnm>S. F.</fnm></au>
  </aug>
  <source>\mnras</source>
  <pubdate>2016</pubdate>
  <volume>457</volume>
  <fpage>313</fpage>
  <lpage>319</lpage>
</bibl>

<bibl id="B42">
  <title><p>{The signatures of the parental cluster on field planetary
  systems}</p></title>
  <aug>
    <au><snm>{Cai}</snm><fnm>M. X.</fnm></au>
    <au><snm>{Portegies Zwart}</snm><fnm>S.</fnm></au>
    <au><snm>{van Elteren}</snm><fnm>A.</fnm></au>
  </aug>
  <source>\mnras</source>
  <pubdate>2018</pubdate>
  <volume>474</volume>
  <fpage>5114</fpage>
  <lpage>5121</lpage>
</bibl>

<bibl id="B43">
  <title><p>{The formation of solar-system analogs in young star
  clusters}</p></title>
  <aug>
    <au><snm>{Portegies Zwart}</snm><fnm>S.</fnm></au>
  </aug>
  <source>\aap</source>
  <pubdate>2019</pubdate>
  <volume>622</volume>
  <fpage>A69</fpage>
</bibl>

<bibl id="B44">
  <title><p>{How Do Disks and Planetary Systems in High-mass Open Clusters
  Differ from Those around Field Stars?}</p></title>
  <aug>
    <au><snm>{Vincke}</snm><fnm>K</fnm></au>
    <au><snm>{Pfalzner}</snm><fnm>S</fnm></au>
  </aug>
  <source>\apj</source>
  <pubdate>2018</pubdate>
  <volume>868</volume>
  <fpage>1</fpage>
</bibl>

<bibl id="B45">
  <title><p>{The viscous evolution of circumstellar discs in young star
  clusters}</p></title>
  <aug>
    <au><snm>{Concha-Ram{\'{\i}}rez}</snm><fnm>F.</fnm></au>
    <au><snm>{Vaher}</snm><fnm>E.</fnm></au>
    <au><snm>{Portegies Zwart}</snm><fnm>S.</fnm></au>
  </aug>
  <source>\mnras</source>
  <pubdate>2019</pubdate>
  <volume>482</volume>
  <fpage>732</fpage>
  <lpage>742</lpage>
</bibl>

<bibl id="B46">
  <title><p>{Induced planet formation in stellar clusters: a parameter study of
  star-disc encounters}</p></title>
  <aug>
    <au><snm>{Thies}</snm><fnm>I.</fnm></au>
    <au><snm>{Kroupa}</snm><fnm>P.</fnm></au>
    <au><snm>{Theis}</snm><fnm>C.</fnm></au>
  </aug>
  <source>\mnras</source>
  <pubdate>2005</pubdate>
  <volume>364</volume>
  <fpage>961</fpage>
  <lpage>-970</lpage>
</bibl>

<bibl id="B47">
  <title><p>{How Sedna and family were captured in a close encounter with a
  solar sibling}</p></title>
  <aug>
    <au><snm>{J{\'i}lkov{\'a}}</snm><fnm>L.</fnm></au>
    <au><snm>{Portegies Zwart}</snm><fnm>S.</fnm></au>
    <au><snm>{Pijloo}</snm><fnm>T.</fnm></au>
    <au><snm>{Hammer}</snm><fnm>M.</fnm></au>
  </aug>
  <source>\mnras</source>
  <pubdate>2015</pubdate>
  <volume>453</volume>
  <fpage>3157</fpage>
  <lpage>-3162</lpage>
</bibl>

<bibl id="B48">
  <title><p>{Is there an exoplanet in the Solar system?}</p></title>
  <aug>
    <au><snm>{Mustill}</snm><fnm>A. J.</fnm></au>
    <au><snm>{Raymond}</snm><fnm>S. N.</fnm></au>
    <au><snm>{Davies}</snm><fnm>M. B.</fnm></au>
  </aug>
  <source>\mnras</source>
  <pubdate>2016</pubdate>
  <volume>460</volume>
  <fpage>L109</fpage>
  <lpage>L113</lpage>
</bibl>

<bibl id="B49">
  <title><p>{Did a stellar fly-by shape the planetary system around Pr 0211 in
  the cluster M44?}</p></title>
  <aug>
    <au><snm>{Pfalzner}</snm><fnm>S</fnm></au>
    <au><snm>{Bhandare}</snm><fnm>A</fnm></au>
    <au><snm>{Vincke}</snm><fnm>K</fnm></au>
  </aug>
  <source>\aap</source>
  <pubdate>2018</pubdate>
  <volume>610</volume>
  <fpage>A33</fpage>
</bibl>

<bibl id="B50">
  <title><p>{Encounter-triggered Disk Mass Loss in the Orion Nebula
  Cluster}</p></title>
  <aug>
    <au><snm>{Olczak}</snm><fnm>C.</fnm></au>
    <au><snm>{Pfalzner}</snm><fnm>S.</fnm></au>
    <au><snm>{Spurzem}</snm><fnm>R.</fnm></au>
  </aug>
  <source>\apj</source>
  <pubdate>2006</pubdate>
  <volume>642</volume>
  <fpage>1140</fpage>
  <lpage>-1151</lpage>
</bibl>

<bibl id="B51">
  <title><p>{Can habitable planets form in clustered environments?}</p></title>
  <aug>
    <au><snm>{de Juan Ovelar}</snm><fnm>M.</fnm></au>
    <au><snm>{Kruijssen}</snm><fnm>J. M. D.</fnm></au>
    <au><snm>{Bressert}</snm><fnm>E.</fnm></au>
    <au><snm>{Testi}</snm><fnm>L.</fnm></au>
    <au><snm>{Bastian}</snm><fnm>N.</fnm></au>
    <au><snm>{C{\'a}novas}</snm><fnm>H.</fnm></au>
  </aug>
  <source>\aap</source>
  <pubdate>2012</pubdate>
  <volume>546</volume>
  <fpage>L1</fpage>
</bibl>

<bibl id="B52">
  <title><p>{Accretion discs in astrophysics}</p></title>
  <aug>
    <au><snm>{Pringle}</snm><fnm>J. E.</fnm></au>
  </aug>
  <source>\araa</source>
  <pubdate>1981</pubdate>
  <volume>19</volume>
  <fpage>137</fpage>
  <lpage>-162</lpage>
</bibl>

<bibl id="B53">
  <title><p>{Spiral Arms in Accretion Disk Encounters}</p></title>
  <aug>
    <au><snm>{Pfalzner}</snm><fnm>S.</fnm></au>
  </aug>
  <source>\apj</source>
  <pubdate>2003</pubdate>
  <volume>592</volume>
  <fpage>986</fpage>
  <lpage>-1001</lpage>
</bibl>

<bibl id="B54">
  <title><p>{The Mass Dependence between Protoplanetary Disks and their Stellar
  Hosts}</p></title>
  <aug>
    <au><snm>{Andrews}</snm><fnm>S. M.</fnm></au>
    <au><snm>{Rosenfeld}</snm><fnm>K. A.</fnm></au>
    <au><snm>{Kraus}</snm><fnm>A. L.</fnm></au>
    <au><snm>{Wilner}</snm><fnm>D. J.</fnm></au>
  </aug>
  <source>\apj</source>
  <pubdate>2013</pubdate>
  <volume>771</volume>
  <fpage>129</fpage>
</bibl>

<bibl id="B55">
  <title><p>{The three-body problem}</p></title>
  <aug>
    <au><snm>{Musielak}</snm><fnm>Z. E.</fnm></au>
    <au><snm>{Quarles}</snm><fnm>B.</fnm></au>
  </aug>
  <source>Reports on Progress in Physics</source>
  <pubdate>2014</pubdate>
  <volume>77</volume>
  <issue>6</issue>
  <fpage>065901</fpage>
</bibl>

<bibl id="B56">
  <title><p>{Disk-Disk Encounters between Low-Mass Protoplanetary Accretion
  Disks}</p></title>
  <aug>
    <au><snm>{Pfalzner}</snm><fnm>S.</fnm></au>
    <au><snm>{Umbreit}</snm><fnm>S.</fnm></au>
    <au><snm>{Henning}</snm><fnm>T.</fnm></au>
  </aug>
  <source>\apj</source>
  <pubdate>2005</pubdate>
  <volume>629</volume>
  <fpage>526</fpage>
  <lpage>-534</lpage>
</bibl>

<bibl id="B57">
  <title><p>{The relation between the most-massive star and its parental star
  cluster mass}</p></title>
  <aug>
    <au><snm>{Weidner}</snm><fnm>C.</fnm></au>
    <au><snm>{Kroupa}</snm><fnm>P.</fnm></au>
    <au><snm>{Bonnell}</snm><fnm>I. A. D.</fnm></au>
  </aug>
  <source>\mnras</source>
  <pubdate>2010</pubdate>
  <volume>401</volume>
  <fpage>275</fpage>
  <lpage>-293</lpage>
</bibl>

<bibl id="B58">
  <title><p>{The Evolution of Protoplanetary Disks in the Arches
  Cluster}</p></title>
  <aug>
    <au><snm>{Olczak}</snm><fnm>C.</fnm></au>
    <au><snm>{Kaczmarek}</snm><fnm>T.</fnm></au>
    <au><snm>{Harfst}</snm><fnm>S.</fnm></au>
    <au><snm>{Pfalzner}</snm><fnm>S.</fnm></au>
    <au><snm>{Portegies Zwart}</snm><fnm>S.</fnm></au>
  </aug>
  <source>\apj</source>
  <pubdate>2012</pubdate>
  <volume>756</volume>
  <issue>2</issue>
  <fpage>123</fpage>
</bibl>

<bibl id="B59">
  <title><p>{Disc-mass distribution in star-disc encounters}</p></title>
  <aug>
    <au><snm>{Steinhausen}</snm><fnm>M.</fnm></au>
    <au><snm>{Olczak}</snm><fnm>C.</fnm></au>
    <au><snm>{Pfalzner}</snm><fnm>S.</fnm></au>
  </aug>
  <source>\aap</source>
  <pubdate>2012</pubdate>
  <volume>538</volume>
  <fpage>A10</fpage>
</bibl>

<bibl id="B60">
  <title><p>{Numerical simulations of protostellar encounters - I. Star-disc
  encounters}</p></title>
  <aug>
    <au><snm>{Boffin}</snm><fnm>H. M. J.</fnm></au>
    <au><snm>{Watkins}</snm><fnm>S. J.</fnm></au>
    <au><snm>{Bhattal}</snm><fnm>A. S.</fnm></au>
    <au><snm>{Francis}</snm><fnm>N.</fnm></au>
    <au><snm>{Whitworth}</snm><fnm>A. P.</fnm></au>
  </aug>
  <source>\mnras</source>
  <pubdate>1998</pubdate>
  <volume>300</volume>
  <fpage>1189</fpage>
  <lpage>1204</lpage>
</bibl>

<bibl id="B61">
  <title><p>{Angular Momentum Transfer in Star-Disk Encounters: The Case of
  Low-Mass Disks}</p></title>
  <aug>
    <au><snm>{Pfalzner}</snm><fnm>S.</fnm></au>
  </aug>
  <source>\apj</source>
  <pubdate>2004</pubdate>
  <volume>602</volume>
  <fpage>356</fpage>
  <lpage>-362</lpage>
</bibl>

<bibl id="B62">
  <title><p>{Gravitational instabilities induced by cluster environment? The
  encounter-induced angular momentum transfer in discs}</p></title>
  <aug>
    <au><snm>{Pfalzner}</snm><fnm>S.</fnm></au>
    <au><snm>{Olczak}</snm><fnm>C.</fnm></au>
  </aug>
  <source>\aap</source>
  <pubdate>2007</pubdate>
  <volume>462</volume>
  <fpage>193</fpage>
  <lpage>-198</lpage>
</bibl>

<bibl id="B63">
  <title><p>{A tidal encounter caught in the act: modelling a star-disc fly-by
  in the young RW Aurigae system}</p></title>
  <aug>
    <au><snm>{Dai}</snm><fnm>F.</fnm></au>
    <au><snm>{Facchini}</snm><fnm>S.</fnm></au>
    <au><snm>{Clarke}</snm><fnm>C. J.</fnm></au>
    <au><snm>{Haworth}</snm><fnm>T. J.</fnm></au>
  </aug>
  <source>\mnras</source>
  <pubdate>2015</pubdate>
  <volume>449</volume>
  <fpage>1996</fpage>
  <lpage>2009</lpage>
</bibl>

<bibl id="B64">
  <title><p>{Generation of highly inclined protoplanetary discs through single
  stellar flybys}</p></title>
  <aug>
    <au><snm>{Xiang-Gruess}</snm><fnm>M.</fnm></au>
  </aug>
  <source>\mnras</source>
  <pubdate>2016</pubdate>
  <volume>455</volume>
  <fpage>3086</fpage>
  <lpage>-3100</lpage>
</bibl>

<bibl id="B65">
  <title><p>{Outer Solar System Possibly Shaped by a Stellar
  Fly-by}</p></title>
  <aug>
    <au><snm>{Pfalzner}</snm><fnm>S</fnm></au>
    <au><snm>{Bhandare}</snm><fnm>A</fnm></au>
    <au><snm>{Vincke}</snm><fnm>K</fnm></au>
    <au><snm>{Lacerda}</snm><fnm>P</fnm></au>
  </aug>
  <source>\apj</source>
  <pubdate>2018</pubdate>
  <volume>863</volume>
  <fpage>45</fpage>
</bibl>

<bibl id="B66">
  <title><p>{The dynamical fate of planetary systems in young star
  clusters}</p></title>
  <aug>
    <au><snm>{Zheng}</snm><fnm>X</fnm></au>
    <au><snm>{Kouwenhoven}</snm><fnm>M. B. N.</fnm></au>
    <au><snm>{Wang}</snm><fnm>L</fnm></au>
  </aug>
  <source>\mnras</source>
  <pubdate>2015</pubdate>
  <volume>453</volume>
  <issue>3</issue>
  <fpage>2759</fpage>
  <lpage>2770</lpage>
</bibl>

<bibl id="B67">
  <title><p>{Planetary Systems in Star Clusters: the dynamical evolution and
  survival}</p></title>
  <aug>
    <au><snm>{Flammini Dotti}</snm><fnm>F</fnm></au>
    <au><snm>{Cai}</snm><fnm>MX</fnm></au>
    <au><snm>{Spurzem}</snm><fnm>R</fnm></au>
    <au><snm>{Kouwenhoven}</snm><fnm>M. B. N.</fnm></au>
  </aug>
  <source>arXiv e-prints</source>
  <pubdate>2018</pubdate>
  <fpage>arXiv:1811.12660</fpage>
</bibl>

<bibl id="B68">
  <title><p>{Survival rates of planets in open clusters: the Pleiades, Hyades,
  and Praesepe clusters}</p></title>
  <aug>
    <au><snm>{Fujii}</snm><fnm>M. S.</fnm></au>
    <au><snm>{Hori}</snm><fnm>Y.</fnm></au>
  </aug>
  <source>\aap</source>
  <pubdate>2019</pubdate>
  <volume>624</volume>
  <fpage>A110</fpage>
</bibl>

</refgrp>
} 


\section{FIGURES}

\begin{itemize}
\item Figure 1: Encounter scenarios; Encounter orbit with periastron $r_{\mathrm{p}}$ in the disk plane  \mbox{$\omega = 0^{\circ}$} (left) and below the disk plane $\omega = 90^{\circ}$ (right).
\item Figure 2: Database structure; Shown here is the tree describing the structure of the datasets stored in hdf5 file format. The mass ratio is the main group and the rest are sub-groups.
\item Figure 3: Particle properties; Face-on disk plots showing particle eccentricity (top) and particle inclination (bottom) at the final time step after an encounter at periastron distance of 100 au by a 1 $M_{\odot}$ perturber at orbital inclination of $60^{\circ}$ and angle of peristron of $0^{\circ}$. 
\end{itemize}

\end{document}